\documentclass[12pt]{article}

\usepackage{amsmath,amssymb}
\usepackage{pifont,fancybox,shadow}
\usepackage{graphicx}
\usepackage{psfrag}
\usepackage{epsf,rotate}
\usepackage[english]{babel}

\usepackage{epsfig}
\usepackage{rotating}

\begin{document}
\baselineskip = 20pt

\def\gvect#1{\mbox{\boldmath $#1$}}

\newcommand{\TIT}[1]{\begin{center}\shadowbox{\Huge\sc \vbox{#1}}\end{center}}
\newcommand{\heading}[1]{\begin{center}\shadowbox{\Large\bf \vbox{#1}}
\end{center}}


\title{Critical behavior of a one-dimensional
monomer-dimer reaction model with lateral interactions.}

\author{Roberto A. Monetti \\ Institut f\"ur Theoretische Physik \\
Physik-Department der Technischen
Universit\"at M\"unchen \\ James-Franck-Strasse, 85747 Garching, Germany}

\maketitle
\begin{abstract}
A monomer-dimer reaction lattice model with lateral 
repulsion among the same
species is studied using a mean-field analysis and Monte Carlo 
simulations. For weak repulsions, the model exhibits a first-order 
irreversible phase
transition between two absorbing states saturated by each different species. 
Increasing
the repulsion, a reactive stationary state appears in addition to the
saturated states. The irreversible phase transitions from the reactive phase 
to any of the saturated states are continuous and belong to the directed 
percolation universality class. However, a different critical behavior is found
at the point where the directed percolation phase boundaries meet. The values
of the critical exponents 
calculated at the bicritical point are in good agreement with the exponents 
corresponding to the parity-conserving universality class. Since the
adsorption-reaction processes does not lead to a non-trivial local
parity-conserving dynamics, this result confirms that the twofold symmetry 
between absorbing states plays a relevant role in determining the universality
class.  The value of the exponent $\delta_2$, which characterizes the
fluctuations of an interface at the bicritical point, supports the
Bassler-Brown's conjecture which states that this is a new exponent in the
parity-conserving universality class. 
\end{abstract}
\newpage
\section{Introduction}
Nonequilibrium models are relevant to a broad scope of phenomena in physics,
chemistry, biophysics, ecology, etc. A relevant feature of the nonequilibrium 
models which exhibit a second-order irreversible phase transition (IPT) to a 
unique absorbing state
is that their critical behavior is in the directed percolation (DP) 
universality class \cite{Grass1}. DP critical behavior is observed over
wide-ranging problems emerging from different disciplines such as quantum
particle physics \cite{Moshe}, irreversible catalytic systems 
\cite{ZGB,Jens1,Eze}, the contact process \cite{Jens2}, branching annihilating 
random 
walks with an odd number of offspring \cite{BARWO}, etc. This fact led 
Janssen and Grassberger to conjecture that in one-component models continuous
transitions to a single absorbing state are in the DP universality class
\cite{Janss}. Motivated by these findings several models with multiple 
absorbing
states were proposed but no new universality class was found \cite{Grins}. This
proved that a greater number of absorbing states is not enough to obtain a 
critical behavior different from DP. Thus the DP universality class is
apparently extremely robust. 

In contrast to the well-established DP universality class, only a
few exceptions are known that do not belong to this class. The known examples
are models A and B of probabilistic cellular automata \cite{grass2,grass3},
nonequilibrium kinetic Ising models \cite{GOdor}, the interacting monomer-dimer
model with infinite repulsion \cite{KimPark}, and the branching annihilating 
random walks with an even number of offsprings \cite{Dben,Jens3,Jens4}. 
A relevant feature is shared by all these models: the number of particles (or 
kinks) is conserved modulo 2. That is why this new universality class is
sometimes called parity-conserving (PC) class. 

Recently, a hierarchy of unidirectionally coupled DP processes has been 
studied \cite{Uwe1}. It has been shown by means of field-theoretic 
renormalization
group techniques and Monte Carlo simulations that new values of the exponent  
$\beta$ corresponding to the order parameter arise at the multicritical point 
while the dynamical critical exponents take the same values as the ones
corresponding to DP. This new result again poses the question of whether a 
few distinct universality classes are enough to characterize the critical
behavior of nonequilibrium systems \cite{Uwe1}.

Few models that display a phase transition in the PC class where no explicit 
conservation law is present have 
appeared in the literature. The known examples are the generalized 
Domany-Kinzel cellular automata \cite{Haye}, the three-species monomer-monomer
model \cite{BrownBass2} and the monomer-monomer surface reaction model 
\cite{Zhuo,BrownBass1}. 
Since these models do not explicitly conserve any 
quantity modulo 2 they show that rather than parity conservation the symmetry
among absorbing states is the origin for the emergence of a different class 
\cite{Haye,ParkPark}. 

In this work we study the behavior of the one-dimensional monomer-dimer 
surface reaction model with lateral repulsion by means of a mean-field (MF) 
analysis and
Monte Carlo simulations. This model is an extension in one dimension of the
well-known Ziff, Gulari, and Barshad surface-reaction model \cite{ZGB}. 
The model was first proposed by Kim and Park
\cite{KimPark} and studied in the case of infinitely strong repulsions. For
finite repulsions the model has a rich critical behavior displaying first- and
second-order IPT. 

The manuscript is organized
as follows. In section II we begin with a brief description of the model and
show the phase diagram obtained by simulations. In section III we present the
MF analysis and compare MF results with simulation results. 
Section IV contains a detailed analysis of the critical behavior of the model
performed by means of Monte Carlo simulations. In the last section we 
summarize our results.

\section{The Model}
The model we study in this paper was first introduced by Kim and Park 
\cite{KimPark}. The monomer-dimer reaction model with lateral 
interactions can be defined as follows: A monomer $A$ or a dimer $B_2$ adsorb
at the vacant sites of a one-dimensional lattice with probabilities $p$ and $q$
respectively, where $p + q = 1$. Each adsorption attempt begins by selecting 
one
site of the lattice at random and if that site is occupied the trial
ends. Otherwise, if the site is empty, an $A$ ($B_2$) species is selected with
probability $p$ ($q$), respectively. In order to introduce lateral interactions
between the same kind of particles, it is assumed that the adsorption 
probability of the selected species depends on the configuration of the 
adsorbed particles on the nearest
neighboring (NN) sites of the selected one. Then the adsorption probability 
$P_{A}$ can be written as
\begin{equation}
P_{A} = 
\begin{cases}
p & \text{if $\nexists$ NN A} \\
p(1 - r_{A}) & \text{if $\exists$ NN A,}
\end{cases}
\end{equation}
where $0 \leq r_{A} < 1$. For the adsorption
of $B_2$ one has first to select at random a NN site of the empty one and if
that site is occupied the trial ends because the dimer can not be deposited on
the lattice. Otherwise, if the site is empty, the adsorption probability 
$P_{B_{2}}$ is given by
\begin{equation}
P_{B_{2}} = 
\begin{cases}
q & \text{if $\nexists$ NN B} \\
q(1 - r_{B}) & \text{if $\exists$ NN B.}
\end{cases}
\end{equation}
where $0 \leq r_{B} < 1$. In this way, the parameters $r_A$ and
$r_B$ can be interpreted as the repulsive interactions among similar species. 
Unlike species on
adjacent sites react immediately and leave the lattice, leading to a process
limited only by adsorption. 

In this work we study the case $r_A = r_B = r$. Previous studies of this 
model \cite{KimPark,ParkPark} focused on various aspects of the critical 
behavior for
infinite repulsive lateral interactions, i. e. $r_A = r_B = 1$.  In this case,
there are two equivalent absorbing states whose configurations are given
only by monomers occupying the odd or even-numbered lattice sites
respectively. The critical behavior of this model was found to be in the PC 
universality class. \\
Figure \ref{fig1} shows the phase diagram of the system obtained by both 
static and
dynamical Monte Carlo simulations. There are two different absorbing
states characterized by the lattice saturated by $A$ or $B$ species 
respectively. For weak repulsions the
system displays first order IPT between the saturated states, and no
reactive phase is observed. The first order critical points have been 
calculated by
means of static Monte Carlo simulations since correlations at first order IPT
are short ranged. Increasing the repulsion a reactive window appears whose 
edges
are second order critical lines that separate this state from the absorbing
states. The second order IPT have been calculated using dynamic Monte Carlo
simulations since fluctuations in this case are imposing.  \\
For $r = 1$ (infinite repulsion) \cite{KimPark} one can identify since the 
very beginning the
presence of two equivalent absorbing states. Although one can not find a priori
a situation like this for $0 \leq r <1$, a phase point where
both absorbing states are statistically equivalent can exist. This happens 
exactly where both second order lines meet, i. e. at the bicritical point. A
similar phase
diagram was also found in the interacting monomer-monomer reaction model 
\cite{Zhuo}. Very
recently, a careful study at the bicritical point led to the conclusion that
this point is in the PC universality class \cite{BrownBass1}. Given the 
symmetry of the adsorption process in the interacting monomer-monomer 
reaction model the bicritical point was found
on the line $p = q = 0.5$. Since the model we study in this work does not
display this
symmetry, the localization of the bicritical point is more difficult.

\section{Mean-Field Theory}
In order to obtain a qualitative understanding of the behavior of the model a 
mean-field analysis is performed. We consider mean-field approximations 
\cite{Ron} that study the time evolution of blocks up to three sites, 
neglecting higher-order correlations. 

We start considering a one-dimensional system of size $L$ at time $t$. Each
site can be only at three different states, namely, $A$, $B$, or $V$ 
corresponding to
a site occupied by a particle of type $A$, $B$ or empty, respectively. In order
to calculate MF approximations taking into account correlations up to
three lattice sites, we write down the following MF rate equations

\begin{eqnarray}
\label{smfa}
\frac{d \rho_{A}}{dt} &=& p \rho_{VVV} + p(1 - r) (\rho_{AVA} + \rho_{AVV} +
\rho_{VVA}) - q (1 - r) (\rho_{BVVA} + \rho_{AVVB}) \nonumber \\ &&
- q (2 \rho_{AVVA} + \rho_{AVVV} + \rho_{VVVA}) ,
\end{eqnarray}

\begin{eqnarray}
\label{smfb}
\frac{d \rho_{B}}{dt} &=& q(2 \rho_{VVVV} + \rho_{AVVV} + \rho_{VVVA}) + 
q(1 - r) (\rho_{BVVV} + \rho_{VVVB} + \rho_{BVV} + \rho_{VVB}) \nonumber \\ 
&& -p(1 - r) (\rho_{AVB} + \rho_{BVA}) - p(\rho_{BVB} + \rho_{BVV} + 
\rho_{VVB}) ,
\end{eqnarray}

\begin{eqnarray}
\label{pmfa}
\frac{d \rho_{AA}}{dt} &&= p (1 - r) (\rho_{VVA} + \rho_{AVV} + 2 
\rho_{AVA}) 
- q (1 - r) (\rho_{AAVVB} + \rho_{BVVAA})  \nonumber \\ && -q(\rho_{AAVVA} + 
\rho_{AVVAA} + \rho_{AAVVV} + \rho_{VVVAA}) ,
\end{eqnarray}

\begin{eqnarray}
\label{pmfb}
\frac{d \rho_{BB}}{dt} &=& \! q \rho_{VVVV} - p (1 - r) (\rho_{BBVA} + 
\rho_{AVBB}) \nonumber \\ && + q (1 - r) (\rho_{BVVB} + 
\rho_{BVVV} + \rho_{VVVB} + \rho_{BVV} + \rho_{VVB}) \nonumber \\ && 
- p (\rho_{VVBB} + \rho_{BBVV} + \frac{1}{2}  \rho_{BBVB} + 
\frac{1}{2} \rho_{BVBB}) ,
\end{eqnarray}

\begin{eqnarray}
\label{tmfa}
\frac{d \rho_{AAA}}{dt} &=& p (1 - r) (\rho_{VVAA} + \rho_{AAVV} + 
\rho_{AVA} 
+ \rho_{AAVA} + \rho_{AVAA}) \nonumber \\&& -q (1 - r) (\rho_{BVVAAA} + 
\rho_{AAAVVB}) 
\nonumber \\&& -q (\rho_{AAAVVA} + \rho_{AVVAAA} + \rho_{VVVAAA} + 
\rho_{AAAVVV}) ,
\end{eqnarray}

\begin{eqnarray}
\label{tmfb}
\frac{d\rho_{BBB}}{dt} &=& \! q (1 - r) (2 \rho_{BVVB} + \rho_{VVBB} + 
\rho_{BBVV} + \rho_{BVVV} + \rho_{VVVB}) \nonumber \\ && 
- p (1 - r) (\rho_{BBBVA} + \rho_{AVBBB}) \nonumber \\ && 
-p(\rho_{VVBBB} + \rho_{BBBVV} + \frac{1}{2}\rho_{BVBBB} + \frac{1}{2}
\rho_{BBBVB}) ,
\end{eqnarray}
\begin{eqnarray}
\label{tmfava}
\frac{d \rho_{AVA}}{dt} &=& p (\rho_{AVVV} + \rho_{VVVA}) + 2p (1 - r)
\rho_{AVVA} - p (1 - r) \rho_{AVA} \nonumber \\&&  -q (1 - r) (\rho_{BVVAVA} 
+ \rho_{AVAVVB}) \nonumber \\&& -q(\rho_{AVAVVA} + \rho_{AVVAVA} + 
\rho_{VVVAVA} + \rho_{AVAVVV}) ,
\end{eqnarray}

\begin{eqnarray}
\label{tmfbvb}
\frac{d \rho_{BVB}}{dt} &=& q(\rho_{BVVVV} + \rho_{VVVVB} + \rho_{AVVVB} + 
\rho_{BVVVA}) \nonumber \\ && + 2 q (1 - r) 
\rho_{BVVVB} - p (1 - r) 
(\rho_{BVBVA} + \rho_{AVBVB}) \nonumber \\ && -p(\rho_{VVBVB} + 
\rho_{BVBVV} + \rho_{BVBVB} + \rho_{BVB}) ,
\end{eqnarray}

and

\begin{eqnarray}
\label{tmfavb}
\frac{d \rho_{AVB}}{dt} &=& p \rho_{VVVB} + q(\rho_{VVVVA} + \rho_{AVVVA})
+ p (1 - r) \rho_{AVVB} \nonumber \\&& + q (1 - r) \rho_{AVVVB} - 
p (\rho_{AVBVV} + 
\frac{1}{2}\rho_{AVBVB}) - p (1 - r) \rho_{AVBVA} \nonumber \\ && - 
q (\rho_{VVVAVB} + \rho_{AVVAVB}) - q (1 - r) \rho_{BVVAVB} - p (1 - r) 
\rho_{AVB} ,
\end{eqnarray}
where $\rho_{i_{1}i_{2}i_{3}...}$ is the density of the block 
$i_{1}i_{2}i_{3}...$ and we have used the equations that
link the density of an $n$-block and the density of an $n+1$-block 
\begin{equation}
\label{relat}
\rho_{i_{1}i_{2}...i_{n}} = \sum_{i_{n+1}} \rho_{i_{1}i_{2}...i_{n}i_{n+1}} = 
\sum_{i_{0}} \rho_{i_{0}i_{1}i_{2}...i_{n}}
\end{equation}
where $AB$ pairs are not allowed since they immediately reacts. The 
processes considered to obtain equations (\ref{smfa}) to (\ref{tmfavb}) are
listed in table 1. 
Furthermore, we obviously have the following constraint
\begin{equation}
\label{const}
\rho_{A} + \rho_{B} + \rho_{V} = 1.
\end{equation}

\begin{center}
\begin{tabular}{||c||c||c||c||} \hline
\multicolumn{4}{|c|}{Table 1: Probabilities for the allowed kinetic 
processes.} \\ 
\hline\hline
$A$ adsorption & Rate & $B_2$ adsorption & Rate \\ \hline
$VVV \rightarrow VAV$ & $p$ & $VVVV \rightarrow VBBV$ & $q$   \\ \hline
$VVA \rightarrow VAA$ & $p(1 - r)$ & $BVVV \rightarrow BBBV$ & $q(1 - r)$ \\ 
\hline
$AVV \rightarrow AAV$ & $p(1 - r)$ & $VVVB \rightarrow VBBB$ & $q(1 - r)$ \\
\hline
$AVA \rightarrow AAA$ & $p(1 - r)$ & $BVVB \rightarrow BBBB$ & $q(1 - r)$ \\
\hline
$VVB \rightarrow VVV$ & $p$ & $AVVV \rightarrow VVBV$ & $q$   \\ \hline
$BVV \rightarrow VVV$ & $p$ & $VVVA \rightarrow VBVV$ & $q$   \\ \hline
$BVB \rightarrow BVV$ & $1/2 \, p$ & $AVVA \rightarrow VVVV$ & $q$   \\ 
\hline
$BVB \rightarrow VVB$ & $1/2 \, p$ & --- & ---   \\ \hline
$AVB \rightarrow AVV$ & $p(1 - r)$ & $AVVB \rightarrow VVBB$ & $q(1 - r)$ \\
\hline
$BVA \rightarrow VVA$ & $p(1 - r)$ & $BVVA \rightarrow BBVV$ & $q(1 - r)$ \\
\hline
\end{tabular}
\end{center}

In the simple MF analysis we just neglect correlations among
sites, i. e. $\rho_{i_{1}i_{2}i_{3}...i_{m}} \approx \rho_{i_{1}}\rho_{i_{2}}
\rho_{i_{3}}... \rho_{i_{m}}$. Within this
approximation, equations (\ref{smfa}), (\ref{smfb}), and (\ref{const}), 
comprise a
close set of equations. Then a solution for $\rho_{A}$, $\rho_{B}$, and 
$\rho_{V}$ can be obtained. However, more equations are needed in order to
obtain results for higher order MF analysis.  
In the pair MF approach we approximate the density of blocks longer 
than two sites in the following way 
\begin{equation}
\rho_{i_{1}i_{2}i_{3}...i_{m}} \approx \rho_{i_{1}i_{2}}
\prod_{j=2}^{m-1}\frac{\rho_{i_{j}i_{j+1}}}{\rho_{i_{j}}}
\end{equation}
Using equations (\ref{smfa})-(\ref{pmfb}), (\ref{const}) and writing down
the equations (\ref{relat}) which relate single-site densities with 
pair densities a solution in the pair-MF approximation can be obtained.
Proceeding in a similar way, within the three-sites MF approximation density of
blocks longer than three sites are replaced by 
\begin{equation}
\rho_{i_{1}i_{2}i_{3}...i_{m}} \approx \rho_{i_{1}i_{2}i_{3}}
\prod_{j=2}^{m-2}\frac{\rho_{i_{j}i_{j+1}i_{j+2}}}{\rho_{i_{j}i_{j+1}}}
\end{equation}
Due to the immediate reaction of $AB$ pairs, the stationary densities of 
triples $AVB$ and $BVA$ are equal. Then, 
by means of equations (\ref{smfa}) to (\ref{tmfavb}), and (\ref{const}), and
considering all the relations between single-site densities, density of pairs
and density of triples, the three-sites MF approximation can be
solved. 
It should be pointed out that due to the complexity of the set of equations 
all the MF approximations have been solved numerically. \\
Figure \ref{fig2} shows plots of the densities of $A$ and $B$ in the stationary
state versus $p$ and $q$ respectively for two different values of the 
repulsion $r$ obtained by static Monte Carlo simulations, simple, pair, and 
three-sites MF analysis. For weak
repulsion ($r = 0.5$) the sharp jump observed in both the densities of $A$ 
and $B$ is the signature of first order IPT. Since correlations at a first
order IPT are finite we expect the MF approaches to give good results. 
Although the simple MF approximation is quite poor, results quickly improve for
the pair and three-sites approaches.
For higher values of the repulsion $r$ the sharp variation in the densities is
no longer present but a smooth transition is observed. In general, MF
theories do not give good results near second order continuous IPT since 
second order IPT
are governed by fluctuations. However, as it is observed in figures 2c and 2d,
we still obtain a fairly good agreement between the three-sites MF
approach and simulation results. \\
Figure \ref{fig3} shows plots of the phase boundary for both the $A$ and $B$ 
saturated phases obtained by simulations, pair, and three-sites MF 
approximations. Within the simple MF analysis the phase
boundary for the $A$ ($B$) saturated phase always occurs at $p = 1$ ($q = 1$),
that is why it is not included in the figure. For weak repulsions, the pair MF
approach gives better results for both the $A$ and $B$ phase boundaries
than the three-sites MF analysis. However, for stronger repulsions,
correlations become more important and the three-sites MF approximation leads 
to better results. It should be pointed out that no bicritical point can be
obtained from the MF approximations considered in this work. However, it is
observed in figure \ref{fig4} that phase boundary curves qualitatively 
resemble the actual phase diagram. 
The closest points
in the phase diagram obtained within the three-site MF approach are 
$(r_{c} \approx 0.6, p_{c}^{A} \approx 0.348)$, $(r_{c} \approx 0.6, p_{c}^{B}
 \approx 0.305)$ which are good approximations for the actual bicritical point
$(r_{c} = 0.559, p_{c} = 0.35)$ (see next section). In the MF treatment of the
interacting
monomer-monomer reaction model \cite{BrownBass1} a bicritical point can always
be found given the symmetry of the adsorption process which is reflected in the
MF rate equations. However, no good approximation for this point was obtained
up to the three-sites MF analysis. 

The monomer-dimer reaction model with lateral interactions displays a feature
that is not present in the interacting monomer-monomer reaction model. In
fact, for $q = 1$, the stationary density $\rho_B$ is always less than one in
spite of the value of the repulsion $r$. However, $\rho_B$ is a function of $r$
since local configurations like $BBVBB$ are more likely to occur when the
repulsion is increased. Then, for $q = 1$, we have a one-dimensional random
dimer filling problem with lateral interactions. In this case 
$\rho_B$ is commonly called jamming coverage which we denote as 
$\Theta_{j}(r)$. \\
Figure \ref{fig5} shows a plot of $\Theta_{j}(r)$ versus $r$ obtained by 
simulations,
pair and three-sites MF approximations. It should be noted that a MF analysis
of the jamming coverage requires at least to take into account correlations up
to pairs of sites. For $r=0$ both MF analysis predict 
$\Theta_{j}(0) = 0.8646...$, reproducing the value derived long ago by Flory
\cite{Fl}. As it has also been observed for the phase boundary curves, the
three-sites (pair) MF approximation gives better results for strong (weak)
repulsions. 

We did not calculate higher order MF approximations since the algebra becomes
much more complicated and the MF approaches presented here provide a fairly
good qualitative understanding of the model.

\section{Simulations Results}
As mentioned in the last section, static Monte Carlo simulations
are suitable to obtain the coordinates of the first-order transition
points. However, second-order IPT are dominated by fluctuations, so in a finite
system and close to the critical point, the stationary states of the reactive
phase can irreversibly evolve into the saturated state (absorbing state). Due
to this circumstance, the precise determination of both critical points and
critical exponents is rather difficult. However, this shortcoming can be 
avoided by performing an epidemic analysis \cite{Grass1}. Within this 
context,
the epidemic analysis is usually called "defect dynamics" simulations.
For this purpose one starts, at $t=0$, with a small block of vacant sites in an
otherwise saturated lattice, i.e. a configuration close to one of the absorbing
states. Then, the time evolution of this block is analyzed by measuring the 
following properties: (i) The average number of vacant sites at time $t$,
$N(t)$, (ii) the survival probability of the block at time $t$, $P(t)$, 
and (iii) the average distance over which the block has spread at time $t$, 
$R(t)$. Finite-size effects are absent because the system is taken large 
enough to avoid the presence of vacant sites at the boundaries. 
For this purpose a lattice of $10^4$ cells is enough. Averages are taken over 
$10^5$ different samples. Near the critical point, the number of vacant sites
is often very small. Then, we improve the efficiency of the algorithm by
keeping a list of vacant sites. Time is incremented by $1/n(t)$, where $n(t)$
is the number of vacant sites at time $t$. Time evolution of blocks are 
monitored up to $t =10^5$. At criticality, the following scaling behavior 
holds:
\begin{equation}
\label{Eta}
N(t) \propto t^{\eta},
\end{equation}
\begin{equation}
\label{Delta}
P(t) \propto t^{-\delta},
\end{equation}
and
\begin{equation}
\label{Zeta}
R(t) \propto t^{z/2}
\end{equation}
where $\delta $, $\eta $ and $z$ are \underline{dynamic} exponents. 

At the bicritical point, it is useful to perform another epidemic analysis 
often called "interface dynamics" simulations. In this case one starts at 
$t = 0$ with a 
minimum width interface between two saturated phases. Since $AB$ pairs
immediately react, there must always be at least a vacancy between two
saturated phases. Then, the epidemic always survives and consequently $\delta =
0$. In addition, a second type of "interface dynamics" simulation can also be
performed. In this case the simulation finishes 
when the interface has collapsed back to its initial width 
\cite{BrownBass2,BrownBass1}. At the bicritical point equations (\ref{Eta}), 
(\ref{Delta}), and (\ref{Zeta}) hold but $P(t)$ must be interpreted as the
probability that the interface has not returned to its minimum width.
Interface dynamics simulations give us information about the competitive
growth of different domains. \\
Figure \ref{fig6} shows log-log plots of $N(t)$, $P(t)$, and $R(t)$ for 
different 
values of $p$ close to the phase boundary between the $B$-saturated
and the reactive phase keeping $r = 0.9$ constant. The straight
line obtained for the three quantities mentioned above at $q = 0.835$ is the 
signature of critical behavior, while slight upward (downward) deviations for 
$q = 0.8325$ $(q = 0.8375)$ indicate supercritical (sub-critical) behavior, 
respectively.
In this way we have determined the critical points along both second order 
phase boundary curves. The analysis at the bicritical point is discussed later.
The spreading or epidemic analysis is a powerful method since the 
error bars for the calculations of the critical points are on the third digit.
The critical exponents obtained at various critical points along the phase
boundary curves have the same values (within error bars) and are in good
agreement with the dynamic critical exponents corresponding to DP which are 
the following:
\begin{equation}
\delta \cong 0.162, \: \eta \cong 0.317, \: z \cong 1.282
\end{equation}

By drawing the second order phase boundary curves one gets the first clue to
the position of the bicritical point. The main property of the system at the 
bicritical point is the symmetry of both saturated phases. This means that at
the bicritical point both absorbing states are statistically equivalent. Then,
defect dynamics simulations at the bicritical point should give
identical results no matter what saturated phase is used to start the
simulation. \\ Figure \ref{fig7}
shows log-log plots of $N(t)$, $P(t)$, and $R(t)$ obtained from defect 
dynamics simulations started using different absorbing states at the 
bicritical point. It is observed that the critical behavior up to time $t 
\approx 10^5$ is governed by the same exponents. Then, our best estimation of 
the value of the 
bicritical point is $(r_{c} = 0.559, p_{c} = 0.35)$. It should be pointed out 
that in the study 
of the interacting monomer-monomer reaction model \cite{BrownBass1} the
localization of the bicritical point was easier because of the symmetry of the
adsorption-reaction processes. \\
We obtain the following values for the dynamical critical exponents at the 
bicritical point   
\begin{equation}
\delta = 0.2910 \pm 0.0002, \: \eta = 0.0034 \pm 0.0003, \: 
z = 1.147 \pm 0.0004
\end{equation}
It should be remarked that the error bars merely indicate the statistical 
errors obtained from regressions. 

The exponent $\beta$ which characterizes the critical behavior of the order
parameter has been obtained directly in static simulations. For the present 
model a good choice for the order parameter is the average density of empty
sites $\rho_{V}$. The behavior of the order parameter close to a critical 
point is given by 
\begin{equation}
\label{beta}
\rho_{V} \sim |\Delta|^{\beta}
\end{equation}
where $\Delta$ is the distance from a point within the reactive phase to the
bicritical point. Figure \ref{fig8} shows a plot of $\rho_{V}$ versus 
$\Delta$ where the
points within the reactive phase belong to a straight line that bisects the
reactive window. Although static simulations are known to be quite inaccurate
to determine $\beta$, we found the reasonable value  $\beta = 0.88 \pm 0.005$.
\\
The set of critical exponents calculated at the bicritical point are in good 
agreement with the ones corresponding to the PC universality class. For defect
dynamics simulations at $r > r_{c}$ and close to the bicritical point a 
crossover is observed where the critical behavior at short times (long times)
is governed by the PC (DP) universality class. \\ 
In the following we present the results of the "interface dynamics"
simulations. Figure \ref{fig9} shows plots of the number of vacancies in the 
interface
$N(t)$ and the average size of the interface $R(t)$ obtained using the first
type of "interface dynamics" simulation (where $\delta = 0$). We average over
$10^5$ independent samples and we find the following value for the 
dynamical critical exponents:
\begin{equation}
\eta_{1} = 0.2840 \pm 0.0002, \: z_{1} = 1.1506 \pm 0.0004
\end{equation}
This values are in good agreement with the corresponding critical exponents 
obtained in other models in the PC universality class 
\cite{KimPark,Dben,Jens3,Haye,BrownBass2}. 
It should be noted that the values of $\eta_{1}$
and $z_{1}$ are similar to the ones obtained from "defect dynamics" 
simulations if we only average over surviving runs. Then the first type of 
"interface dynamics" simulation yields no new information about the dynamics 
of the interface and reflects the full equivalence of the absorbing states at
the bicritical point.

Figure \ref{fig10} shows plots of $N(t)$, $P(t)$, and $R(t)$ obtained
using the second kind of "interface dynamics" simulation. We average over $9$
x $10^6$ independent samples and we find the following values for the critical
exponents:
\begin{equation}
\delta_{2} = 0.7163 \pm 0.0003, \: \eta_{2} = -0.4277 \pm 0.0004, \: 
z_{2} = 1.160 \pm 0.0006
\end{equation}
We first observe that the values of the exponents $z_{1}$ and $z_{2}$ are
similar. The exponent $\eta_{1}$ must be compared with $\delta_{2}
+ \eta_{2}$ which governs the time evolution of the vacant sites averaged only
over the surviving runs. Both exponents are equal within error bars. This is 
the signature of the universal behavior of the critical spreading of the 
interface. The value
of the new independent dynamical exponent $\delta_{2}$ supports the 
Bassler-Browne
conjecture \cite{BrownBass2} which states that $\delta_{2}$ is a new universal
exponent within the PC class. \\
Recently Park and Park \cite{ParkPark} have found that the interacting
monomer-dimer reaction model with infinite repulsive interaction ($r=1$)
supports a kink representation where the total number of kinks is conserved
modulo 2. By including a parity-conserving symmetry breaking field that 
favors one of the
absorbing states, the authors showed that the critical behavior of the model
changes from PC to DP. Then, they concluded that the conservation of the
number of kinks modulo 2 is not
the reason for observing a universality class different from DP but the
symmetry of the absorbing states. 

In contrast to the case with infinite repulsion, the present 
variant of the model involve adsorption-reaction processes in which parity is 
not conserved explicitly. Although by their own nature the number of domain 
walls are conserved modulo $2$, it should be remarked that a non-trivial PC
dynamics requires the creation of at least three kinks per step which is not 
possible in our model. This corroborates that the symmetry
among absorbing states is indeed the only essential property of models in the 
PC class \cite{KimPark,Haye,BrownBass2,BrownBass1}. 

It is interesting to discuss the relation between the present 
model and the branching and annihilating walks models. Static Monte Carlo
simulations close to the bicritical point reveal that stationary 
configurations are formed by large clusters of 
different species that survive a long time. Then, it is clear that the
dynamics of domain walls between clusters of different species will be relevant
at the bicritical point. Following the ideas presented in reference 
\cite{Haye}, at the
bicritical point "walkers" can be identified with domain walls between 
clusters of different species. It should be pointed out that we
have to generalize the concept of walkers. In fact, the "walkers" defined above
are extended objects of fluctuating width. It should be noticed that at
the bicritical point the typical width of a walker is only few lattice
sites. Considering the time evolution of this "walkers" in the long-time regime
and for large lattices, an effective parity-conserving dynamics may be
restored. Concerning the dynamics of these "walkers", it is possible to 
identify an inactive and an active phase. In fact, the line defined by the 
first-order phase transition points, corresponds to the inactive phase for the
dynamics of the "walkers", which ends at the bicritical point. A continuation 
of this line from the bicritical point through the 
reactive phase, would correspond to the active phase for the dynamics (see 
figure 1). In other words, the dynamics of this walkers is
defined within a subspace of the phase diagram in which the statistical weight
of the kinetic processes involving the species $A$ and $B$ is the same.  
Recently, Cardy and T\"auber have developed a systematic theory for the
branching and annihilating random walkers \cite{Uwe2}. The authors have shown 
that in one dimension, fluctuation effects lead to the emergence of a 
non-trivial inactive phase for values of the branching rate $0 \le \sigma < 
\sigma_c$, and a dynamic phase transition at $\sigma_{c} > 0$, in contrast to 
the mean-field result $\sigma_{c} = 0$. However, in two dimensions the theory 
predicts that fluctuation effects generate logarithmic corrections and the 
critical branching rate takes the value predicted by the bare mean-field 
theory ($\sigma_{c} = 0$). Then, the identification of an inactive phase for $0
\le r < r_{c}$ and a dynamic phase transition at $r_{c} > 0$ in the present
model, is in complete agreement with the theoretical results. Recent results 
for a two-dimensional monomer-monomer
surface-reaction model with repulsive lateral interactions \cite{Adri} indicate
that the critical branching rate takes the value $r_c = 0$ 
(mean-field value), also supporting the theoretical results. 

The same ideas can also be applied to other models in the PC universality class
that violate local parity conservation \cite{BrownBass2,BrownBass1}.
\section{Conclusions}
We have studied a monomer-dimer surface reaction lattice model with lateral 
repulsion among same species using a mean-field analysis and Monte Carlo 
simulations. For weak repulsions the model exhibits first-order IPT between 
two phases saturated by different species. Increasing the
interaction a reactive window appears whose edges are second-order critical 
lines that separate this state from the absorbing states. 

We have considered MF approximations that take into account
correlations up to three lattice sites. The stationary density of the species,
the phase boundary curves, and the jamming coverage have been studied within
the MF approaches and the results compared with simulations. For weak
(strong) repulsion the pair (three-sites) MF approximation has led us to 
better 
results than the three-sites (pair) MF approximation for all the quantities
mentioned above. No bicritical point has been obtained from the MF
approximations considered in this work. However, phase boundary curves
calculated within MF approximations qualitatively resemble the actual phase
diagram. Within the three-sites MF approach phase boundary curves come up
closest at $(r_{c} \approx 0.6, p_{c}^{A} \approx 0.348)$, $(r_{c} \approx 
0.6, p_{c}^{B} \approx 0.305)$ which are good estimations for the actual 
bicritical point $(r_{c} = 0.559, p_{c} = 0.35)$. 

The critical behavior of the model has been studied using both static and
dynamical Monte Carlo simulations. IPT between the reactive phase and any of
the saturated phases are in the DP universality class. However, it has been 
found a
critical behavior different from DP at the point where both DP curves meet. The
critical exponents calculated at the bicritical point are in good agreement
with the ones corresponding to the PC class. Since this model does not conserve
explicitly any quantity modulo 2, it corroborates that the twofold symmetry in
the absorbing states is the reason for obtaining a critical behavior different
from DP. It is relevant to mention that the authors of the theory for the
branching and annihilating random walkers \cite{Uwe2} have identified a formal
permutation symmetry at the Hamiltonian level in the case of an even number 
of offsprings. This fact again indicates that symmetry among 
absorbing states plays a relevant role in determining the universality class.

The value of the critical exponent $\delta_2 \cong 0.71$ corresponding to
the probability $P(t)$ that the interface has not collapsed back to its minimum
width is in good agreement with the values obtained in other models 
\cite{BrownBass2,BrownBass1,Hw}. This corroborates the Bassler-Brown conjecture
\cite{BrownBass2} which states that the interfacial fluctuations are an 
additional universal characteristic of models in the PC class. 

{\bf Acknowledgments}:
I would like to thank Uwe T\"{a}uber and Ezequiel Albano for a careful reading
of the manuscript. Interesting discussions with Uwe T\"{a}uber and
Haye Hinrichsen are also acknowledged. 
This work was financially supported by CONICET (Argentina).

\newpage
\begin{figure}
\caption{Phase diagram of the model obtained by means of Monte Carlo
simulations.} 
\label{fig1}
\end{figure}

\begin{figure}
\caption{Plots of $\rho_{A}$ and $\rho_{B}$ versus $p$ and $q$ respectively,
obtained using static Monte Carlo simulations, simple, pair, and three-sites MF
approximations. a) $\rho_{A}$ versus $p$ for $r = 0.5$, b) $\rho_{B}$ versus
$q$ for  $r = 0.5$, c) $\rho_{A}$ versus $p$ for $r = 0.9$, and 
d) $\rho_{B}$ versus $q$ for $r = 0.9$.} 
\label{fig2}
\end{figure}

\begin{figure}
\caption{Phase boundary curves obtained by means of Monte Carlo simulations,
pair, and three-sites MF approximations. a) Boundary of the $A$-saturated phase
, b) Boundary of the $B$-saturated phase} 
\label{fig3}
\end{figure}

\begin{figure}
\caption{Phase boundary curves obtained within the 
pair, and three-sites MF approximations. a) Pair MF approximation
b) Three-sites MF approximation.} 
\label{fig4}
\end{figure}

\begin{figure}
\caption{Plot of the jamming coverage $\Theta_j(r)$ versus $r$ obtained using
Monte Carlo simulations, pair, and three-sites MF approximations.} 
\label{fig5}
\end{figure}

\begin{figure}
\caption{Log-log plots obtained using defect dynamics simulations keeping 
$r = 0.9$ constant. a) $N(t)$ versus $t$ b) $P(t)$ versus $t$ c) $R(t)$ versus
$t$.} 
\label{fig6}
\end{figure}

\begin{figure}
\caption{Log-log plots obtained using defect dynamics simulations initiated
using different absorbing states at the
bicritical point. a) $N(t)$ versus $t$ b) $P(t)$ versus $t$ c) $R(t)$ versus
$t$.} 
\label{fig7}
\end{figure}

\begin{figure}
\caption{Plot of the density of vacant sites $\rho_{V}$ versus the distance
$|\Delta|$ to the bicritical point. A straight line of slope $\beta = 0.90$ has
been included.} 
\label{fig8}
\end{figure}

\begin{figure}
\caption{Log-log plots obtained using the first type of interface dynamics 
simulations ($\delta = 0$) for the following values of the parameters: $(r = 
0.57, p = 0.3484)$ (upper curve), $(r = 0.559, p = 0.35)$ (middle curve) 
$(r = 0.559, p = 0.355)$ (lower curve). a) $N(t)$ versus $t$ b) $R(t)$ versus 
$t$.} 
\label{fig9}
\end{figure}

\begin{figure}
\caption{Log-log plots obtained using the second type of interface dynamics 
simulations for the following values of the parameters: $(r = 0.57, 
p = 0.3484)$ (upper curve), $(r = 0.559, p = 0.35)$ (middle curve) 
$(r = 0.559, p = 0.355)$ (lower curve). a) $N(t)$ versus $t$ b) $P(t)$ versus
$t$ c) $R(t)$ versus $t$. The inset amplifies the last decade in order to
distinguish between the three curves since $R(t)$ is less sensitive to changes
in the parameters.} 
\label{fig10}
\end{figure}

\begin{thebibliography}{}
\bibitem{Grass1} P. Grassberger and A. de la Torre,  Ann. Phys. (New York) {\bf
122}, 373 (1979); P. Grassberger, J. Phys A: {\it Math. Gen.} {\bf 22}, 3673
(1989).
\bibitem{Moshe} M. Moshe, Phys. Rep. C {\bf 37}, 255 (1978).
\bibitem{ZGB} R. Ziff, E. Gulari, and Y. Barshad, Phys. Rev. Lett. {\bf 56}, 
2553 (1986).
\bibitem{Jens1} I. Jensen, H. Fogedby, and R. Dickman, Phys. Rev. A {\bf 41},
R3411 (1990).
\bibitem{Eze} E. V. Albano, Phys. Rev. Lett. {\bf 69}, 656 (1992).
\bibitem{Jens2} I. Jensen, Phys. Rev. Lett. {\bf 70}, 1465 (1993).
\bibitem{BARWO} H. Takayashu and A. Y. Tretyakov, Phys. Rev. Lett. {\bf 68},
3060 (1992); I. Jensen, Phys. Rev. E {\bf 47}, 1 (1993).
\bibitem{Janss} H. K. Janssen,  Z. Phys. B {\bf 42}, 151 (1981); P. 
Grassberger, Z. Phys. B {\bf 47}, 365 (1982).
\bibitem{Grins} G. Grinstein, Z. W. Lai  and D. Browne, Phys. Rev. A {\bf 40},
4820 (1989); I. Jensen and R. Dickman, Phys. Rev. E {\bf 48}, 1710
(1993); I. Jensen,  J. Phys. A: {\it Math. Gen.} {\bf 27}, L61 (1994).
\bibitem{grass2} P. Grassberger, F. Krause and T. von der Twer, J. Phys. A 
{\it Math. Gen.} {\bf 17}, L105 (1984).
\bibitem{grass3} P. Grassberger, J. Phys. A {\it Math. Gen.} {\bf 22}, L1103 
(1984).
\bibitem{GOdor} N. Menyh\'{a}rd, J. Phys. A {\it Math. Gen.} {\bf 27}, 6139 
(1994); N. Menyh\'{a}rd and G. \'{O}dor, J. Phys. A {\it Math. Gen.} {\bf 29},
7739 (1996).
\bibitem{KimPark} M. H. Kim and H. Park, Phys. Rev. Lett. {\bf 73},
2579 (1994); H. Park, M. H. Kim, and H. Park, Phys. Rev. E {\bf 52},
5664 (1995); M. H. Kim and H. Park, J. Korean Phys. Soc. {\bf 26},
S345 (1993).
\bibitem{Dben} D. ben-Avraham, F. Leyvraz, and S. Redner, 
Phys Rev. E {\bf 50}, 1843 (1994).
\bibitem{Jens3} I. Jensen, J. Phys. A {\it Math. Gen.} {\bf 26}, 3921 (1993).
\bibitem{Jens4} I. Jensen, Phys. Rev. E {\bf 50}, 3623 (1994).
\bibitem{Uwe1} U. T\"{a}uber, M. Howard, and H. Hinrichsen, cond-matt 9709057,
(1997); U. Alon, M. R. Evans, H. Hinrichsen, and D. Mukamel, Phys. Rev. Lett. 
{\bf 76}, 2746 (1996).
\bibitem{Haye} H. Hinrichsen, Phys. Rev E {\bf 55}, 219 (1997).
\bibitem{BrownBass2} K. E. Bassler and D. A. Brown, Phys. Rev. Lett. {\bf 77},
4094 (1996); Phys. Rev. E {\bf 55}, 5225 (1997).
\bibitem{Zhuo} J. Zhuo, S. Redner, and H. Park, J. Phys. A: {\it Math. Gen.} 
{\bf 26}, 4197 (1993).
\bibitem{BrownBass1} K. S. Brown, K. E. Bassler and D. A. Brown, Phys. Rev E
{\bf 56}, 3953 (1997).
\bibitem{ParkPark} H. Park and H. Park, Physica A {\bf 221}, 97
(1995). 
\bibitem{Ron} R. Dickman, Phys. Rev. A {\bf 34}, 4246 (1986), H. Gutowitz,
J. D. Victor, and B. W. Knight, Physica 28D, 18 (1987).
\bibitem{Fl} P. J. Flory, J. Am. Chem. Soc. {\bf 61}, 1518 (1939).
\bibitem{Uwe2} J. L. Cardy and U. C. T\"auber, Phys. Rev. Lett. {\bf 77}, 4780
(1996); J. L. Cardy and U. C. T\"auber, J. Stat. Phys. {\bf 90}, 1 (1998).
\bibitem{Adri} R. A. Monetti, in preparation.
\bibitem{Hw} W. M. Hwang, S. Kwon, H. Park, and H. Park,
cond-mat/9712259, (1997).
\end{thebibliography}
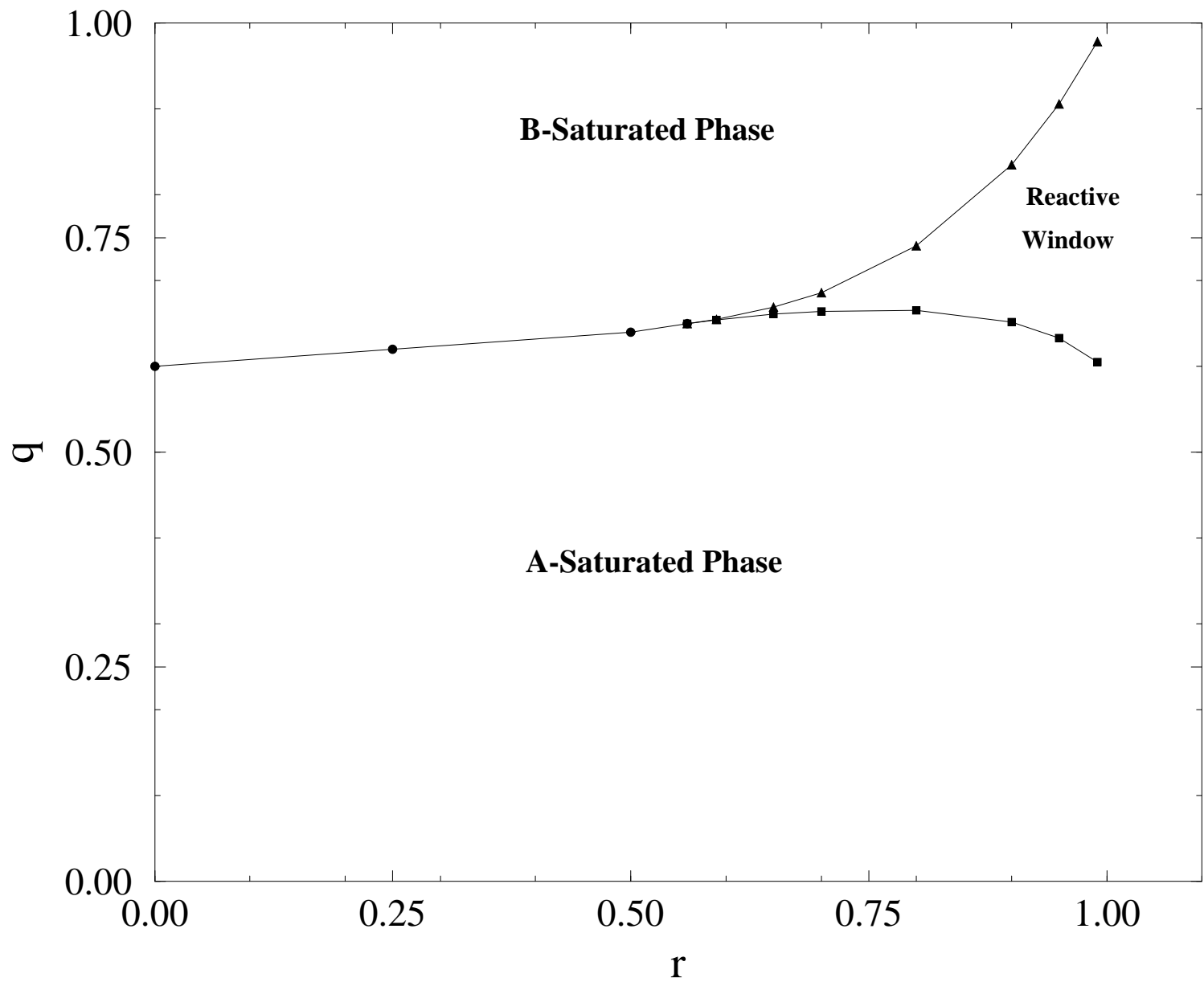
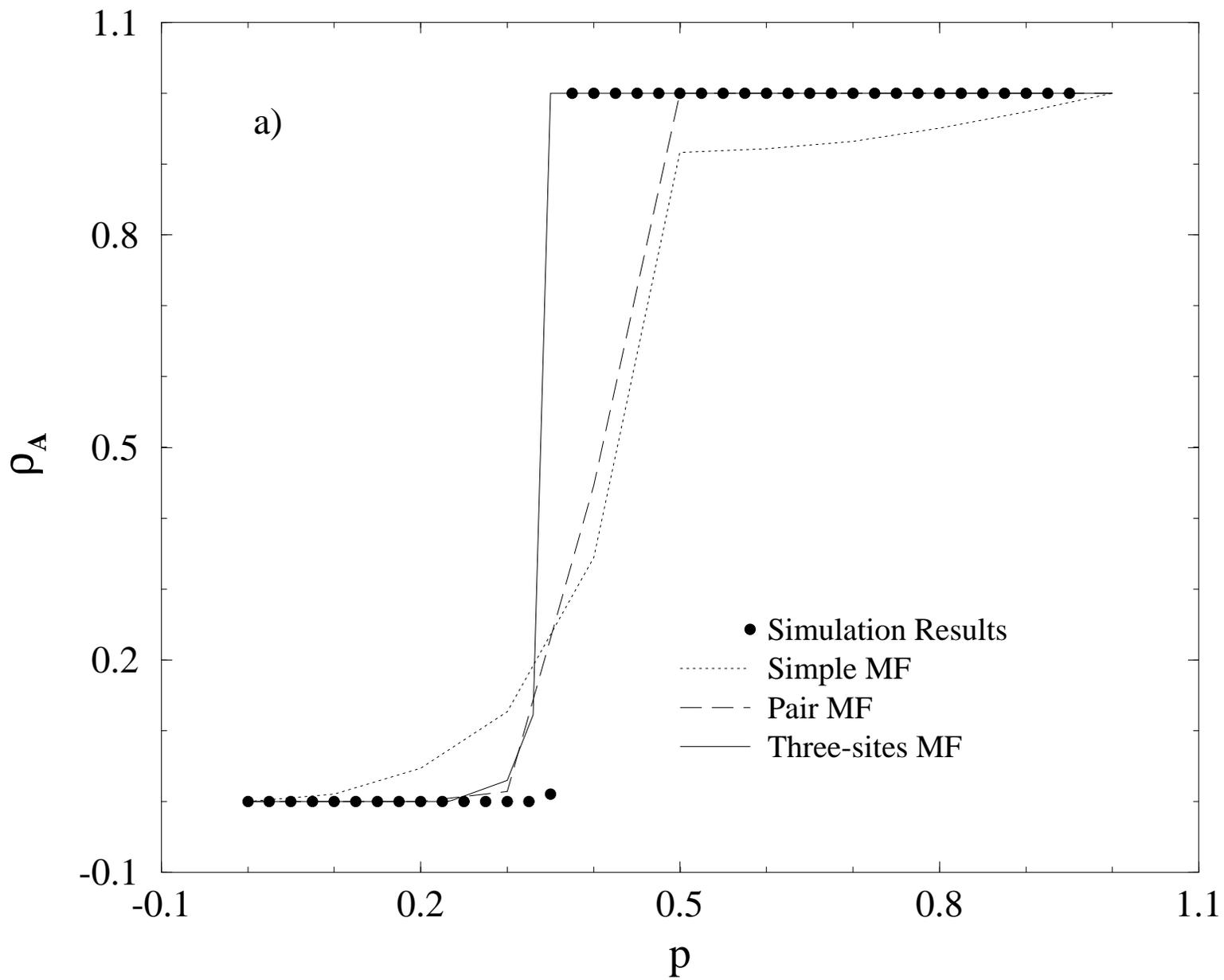
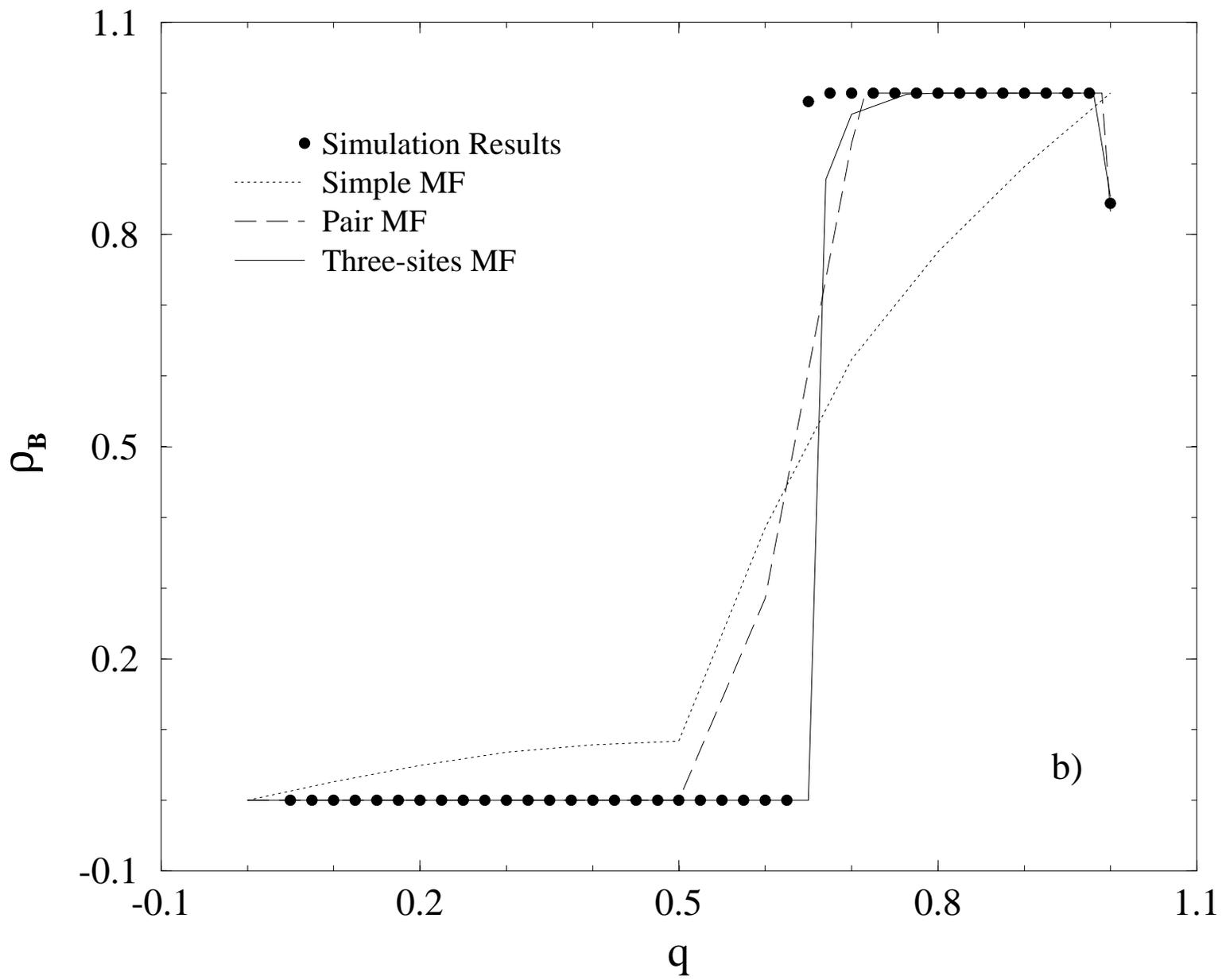
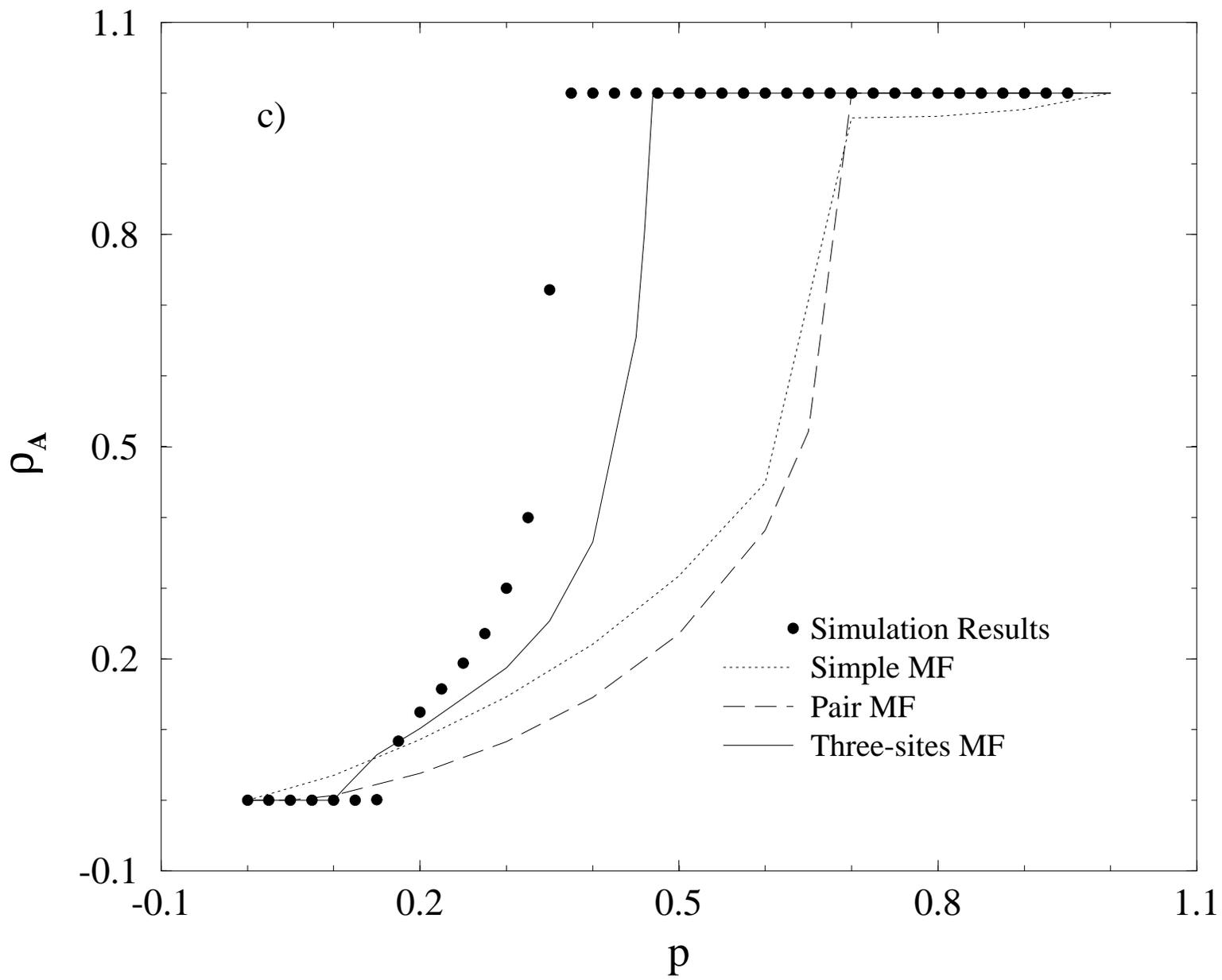
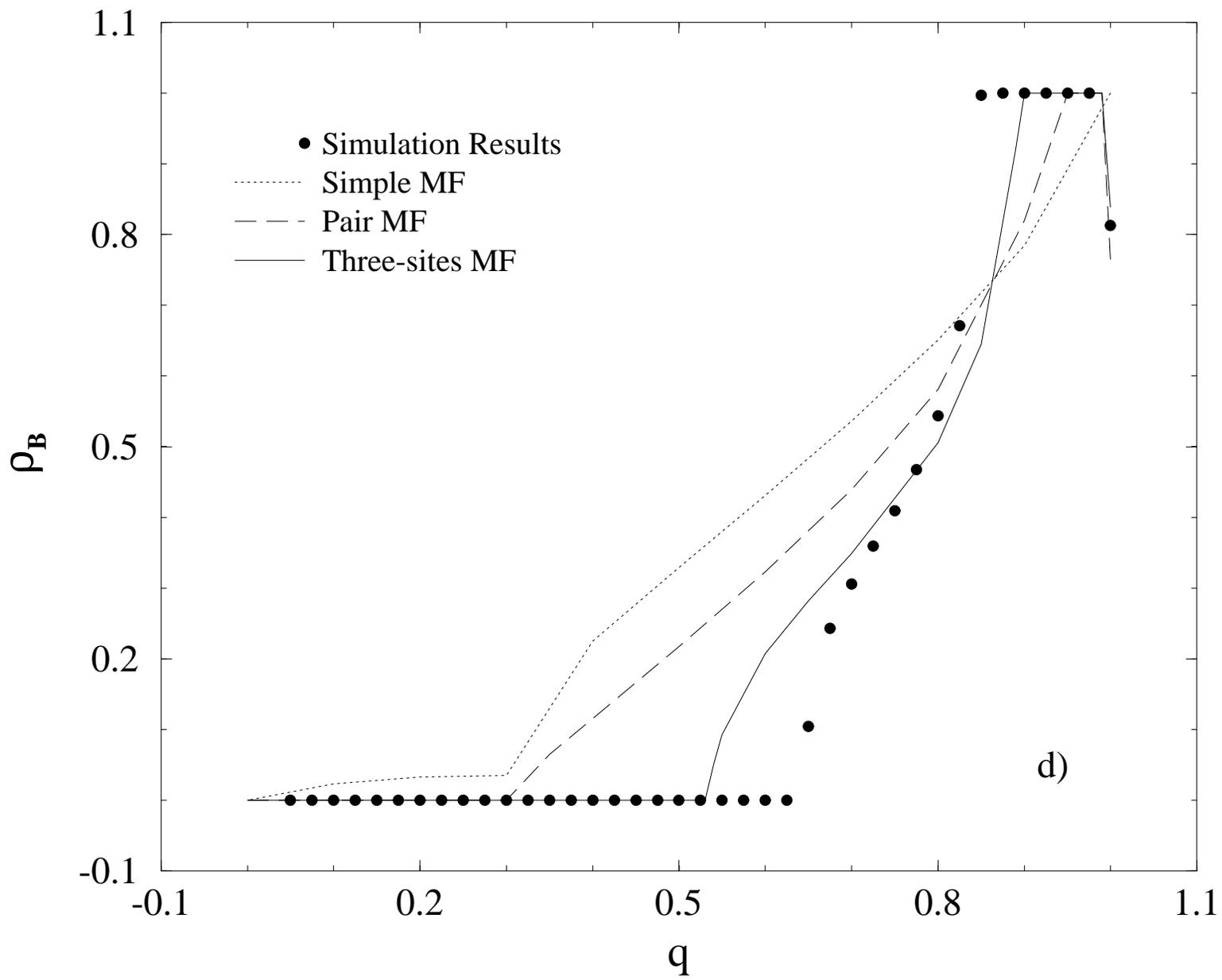
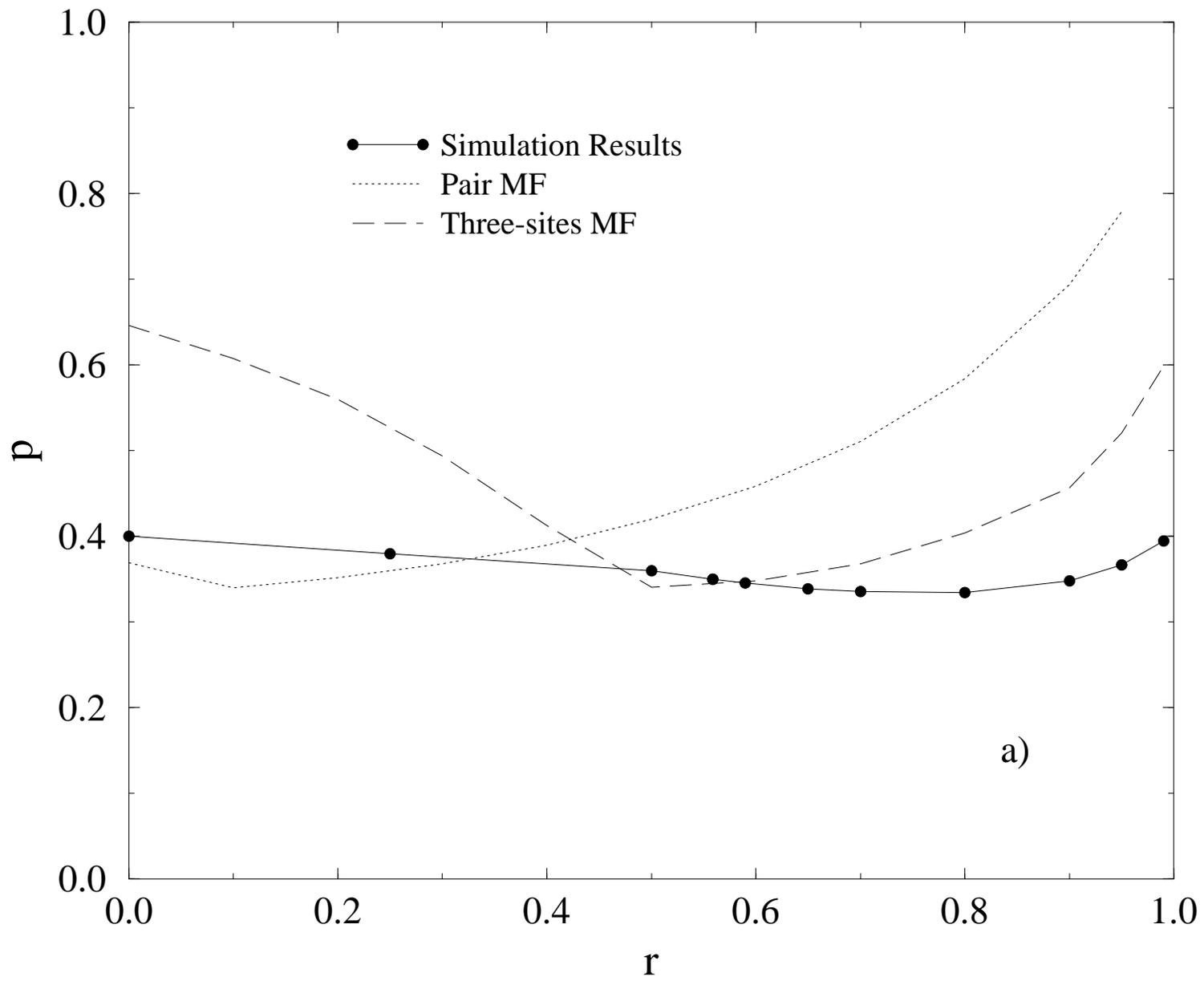
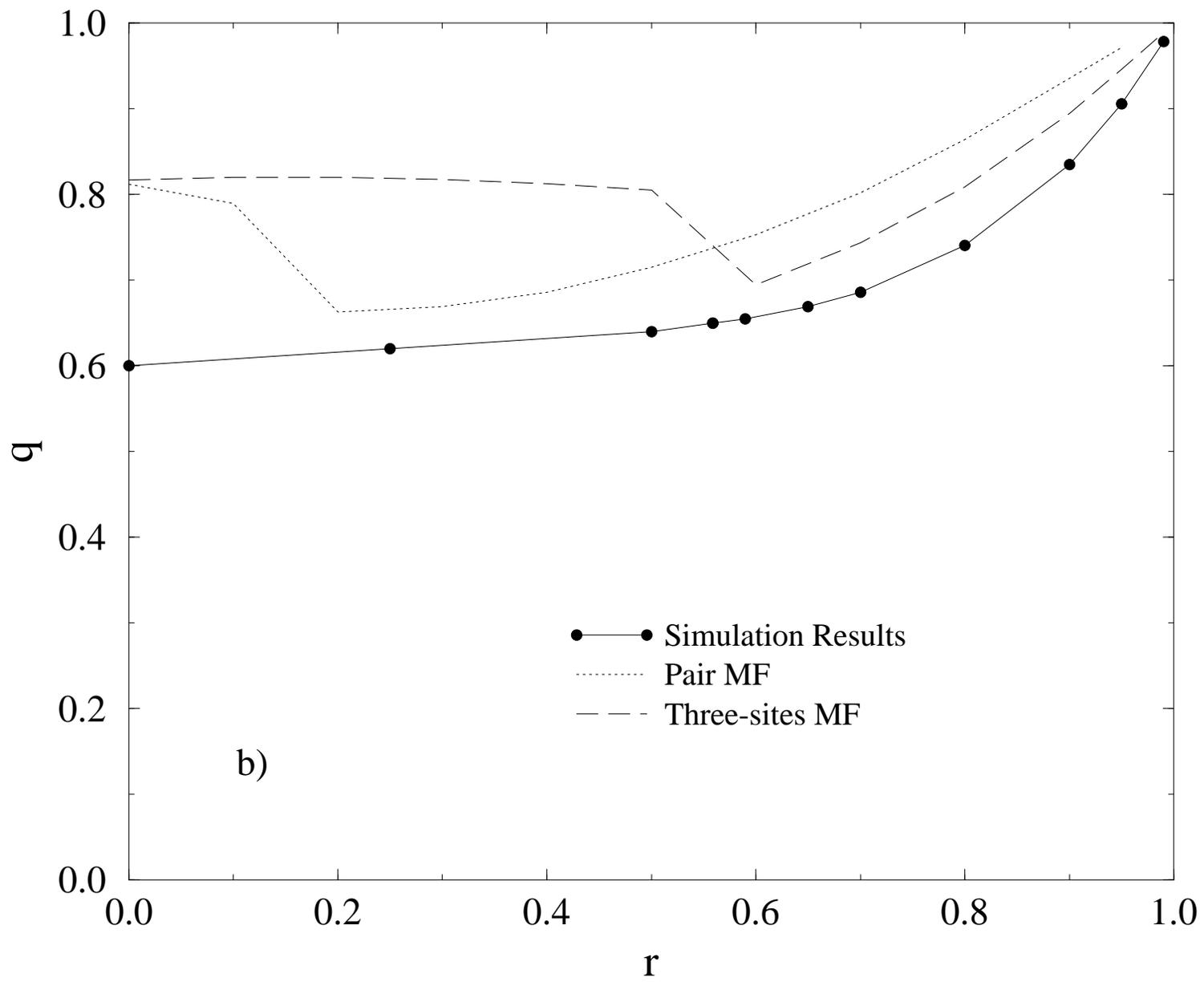
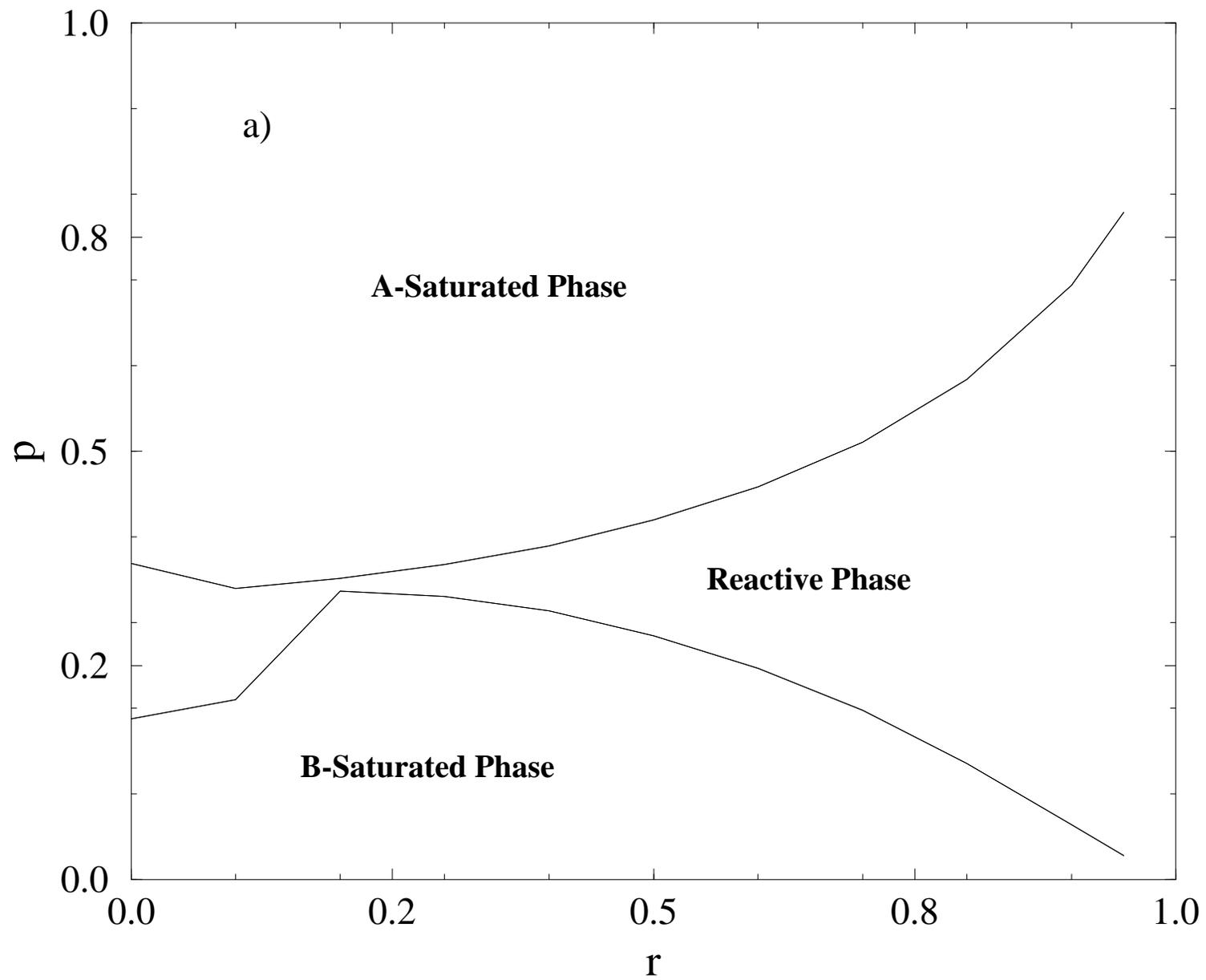
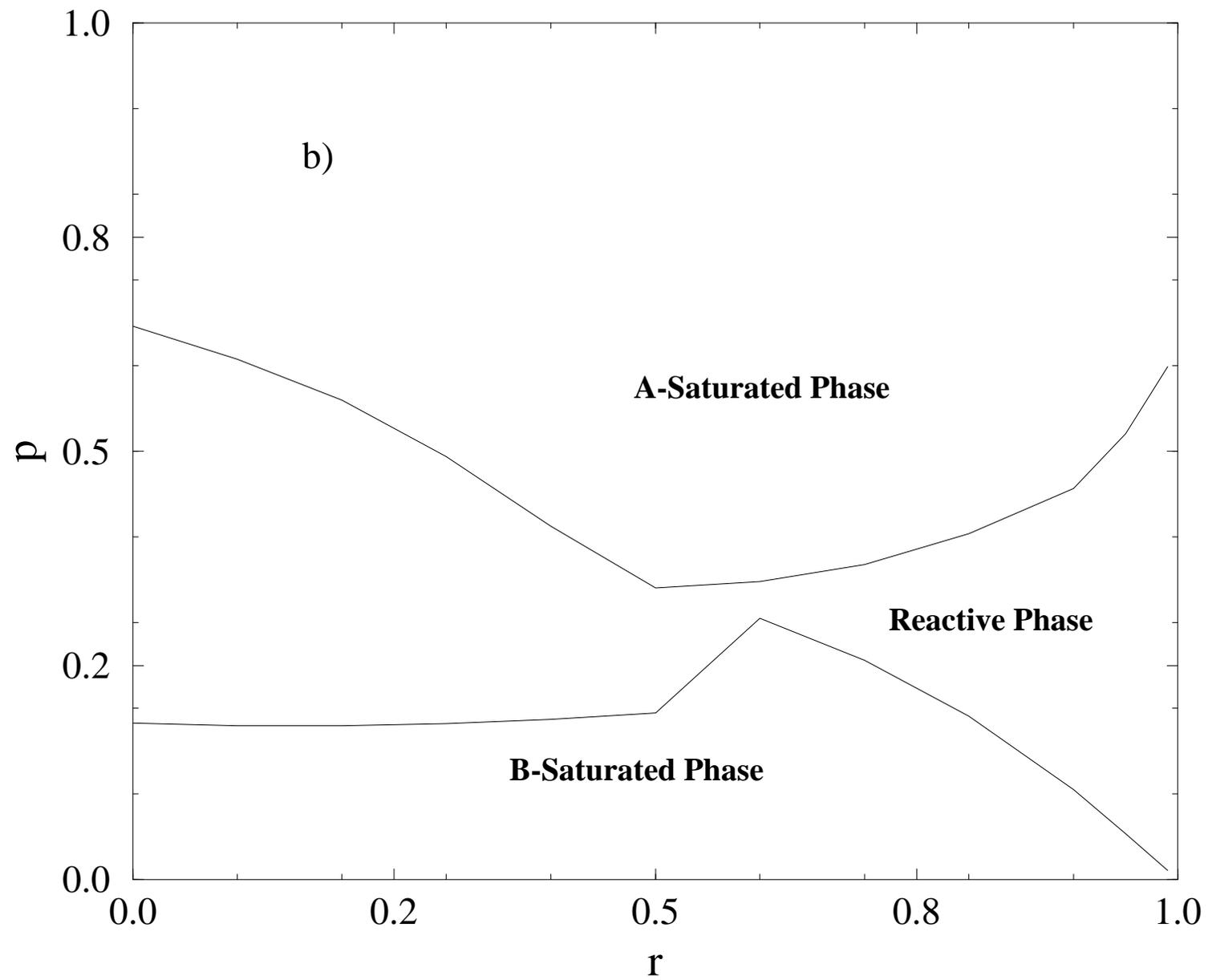
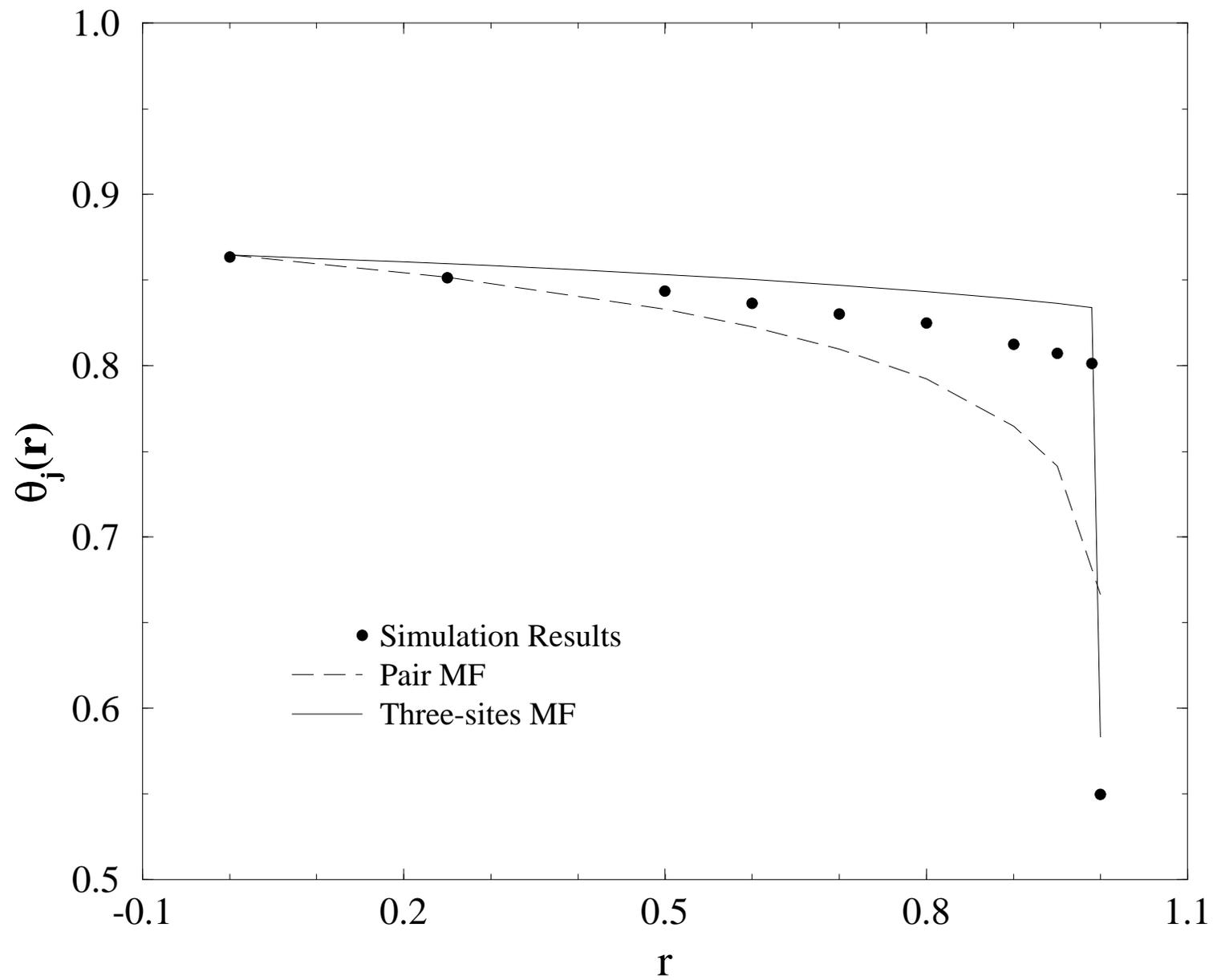
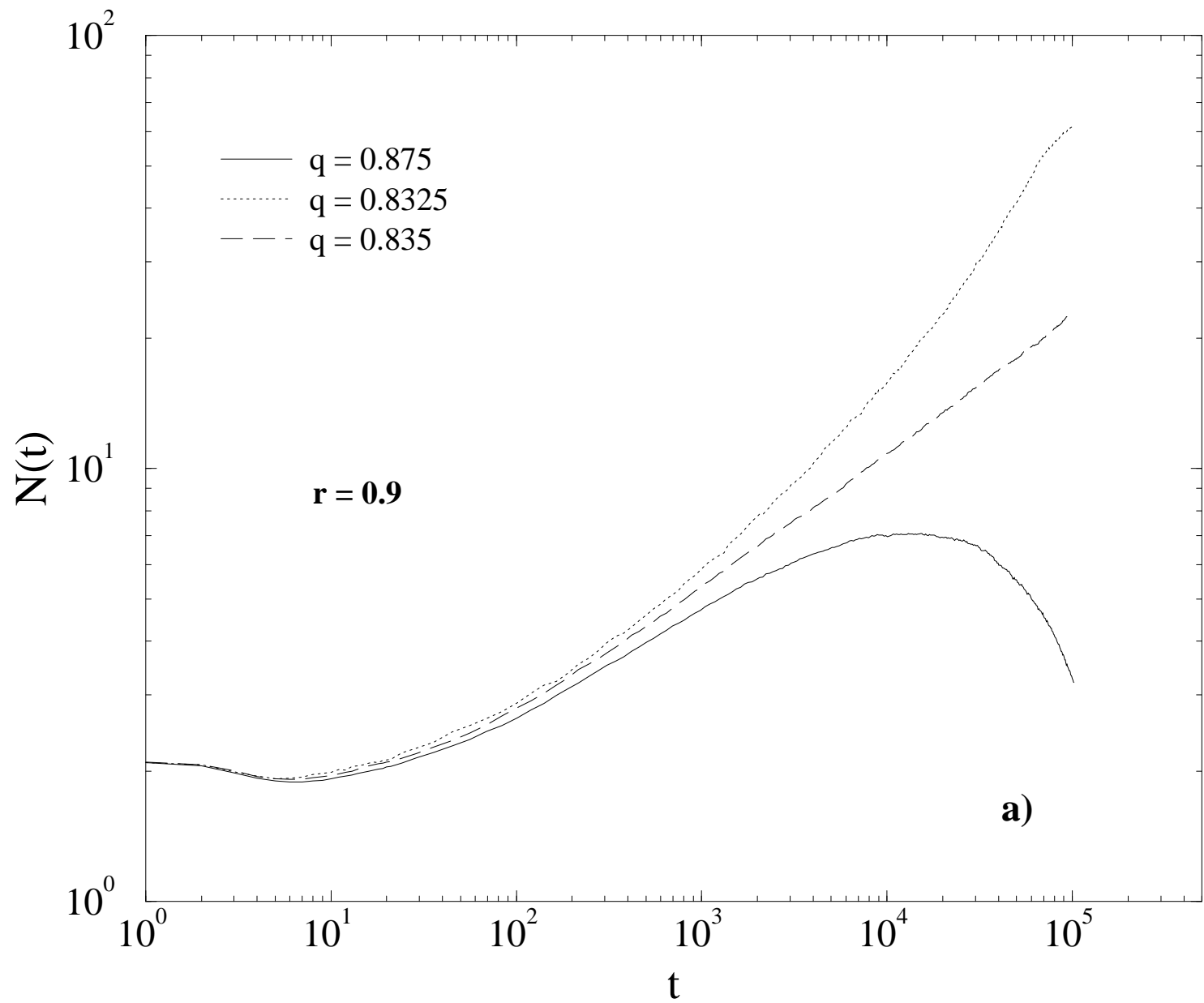
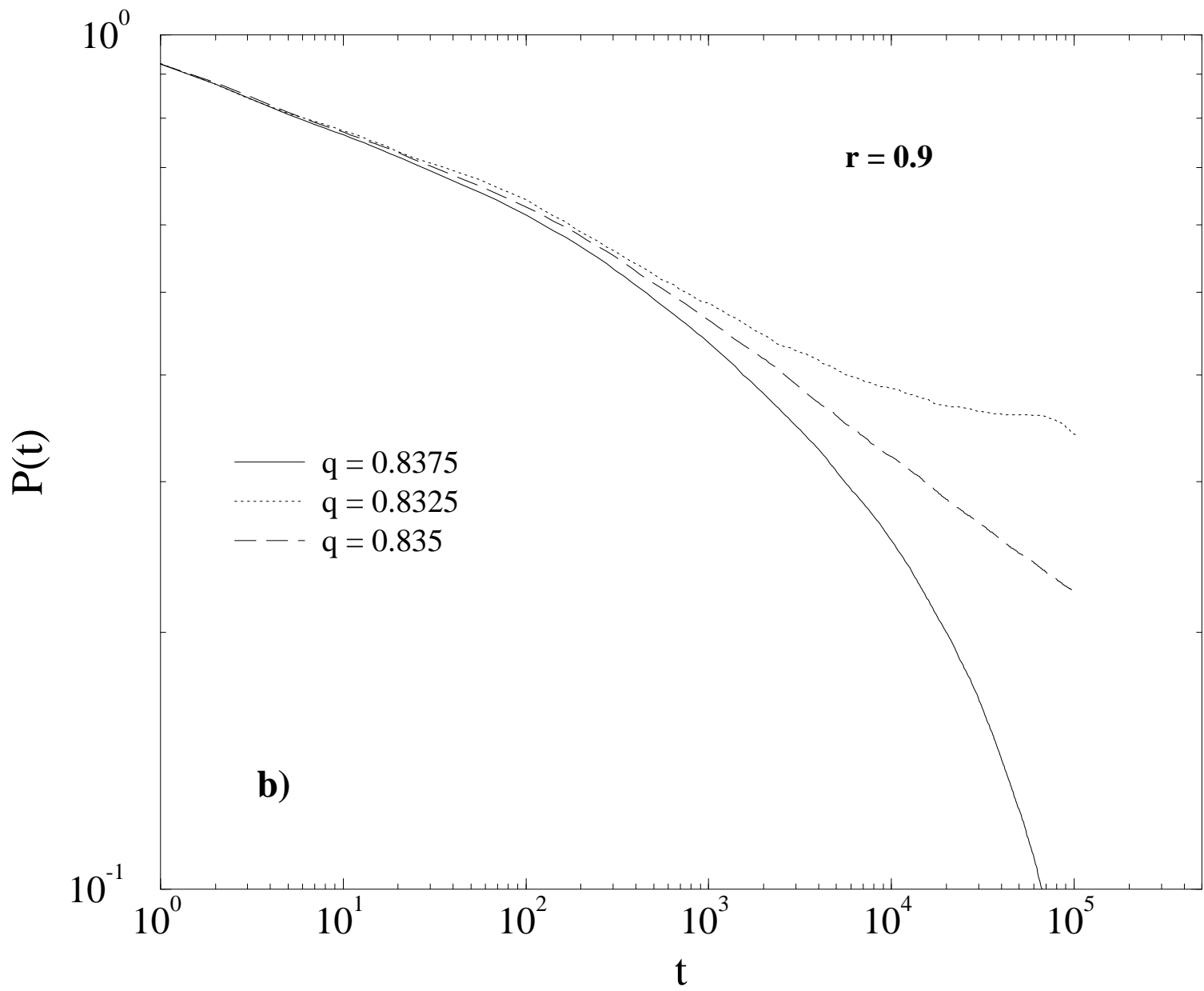
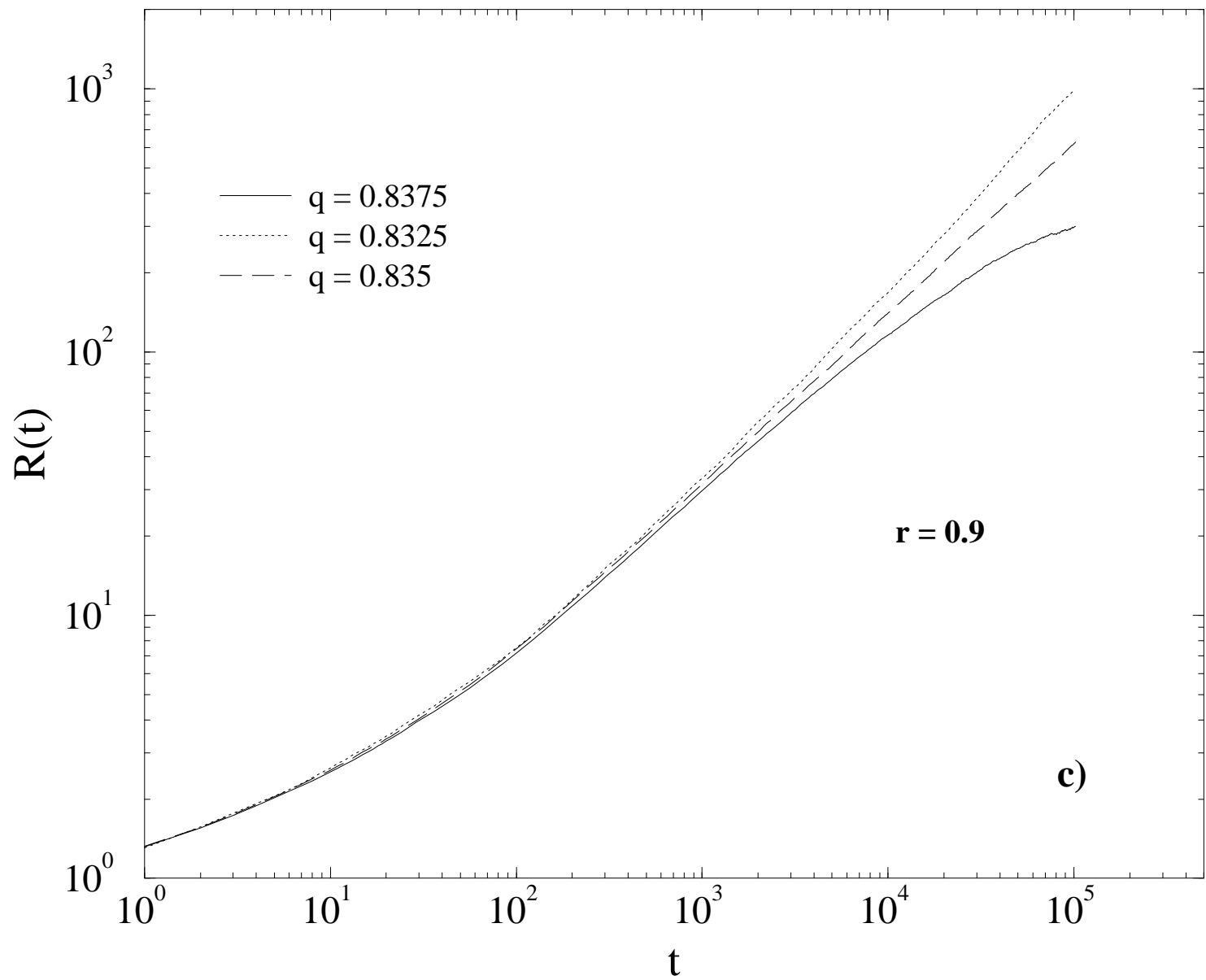
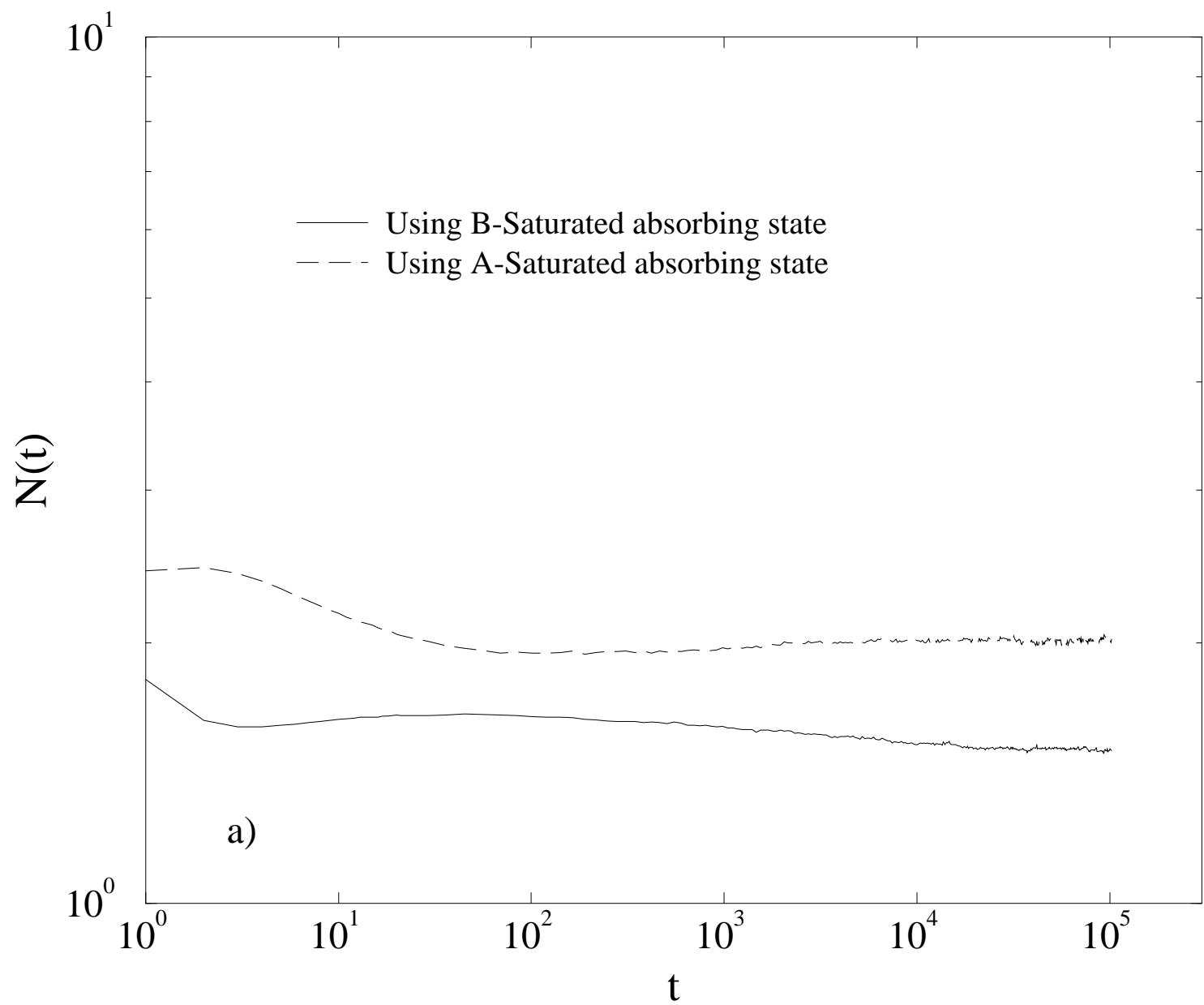
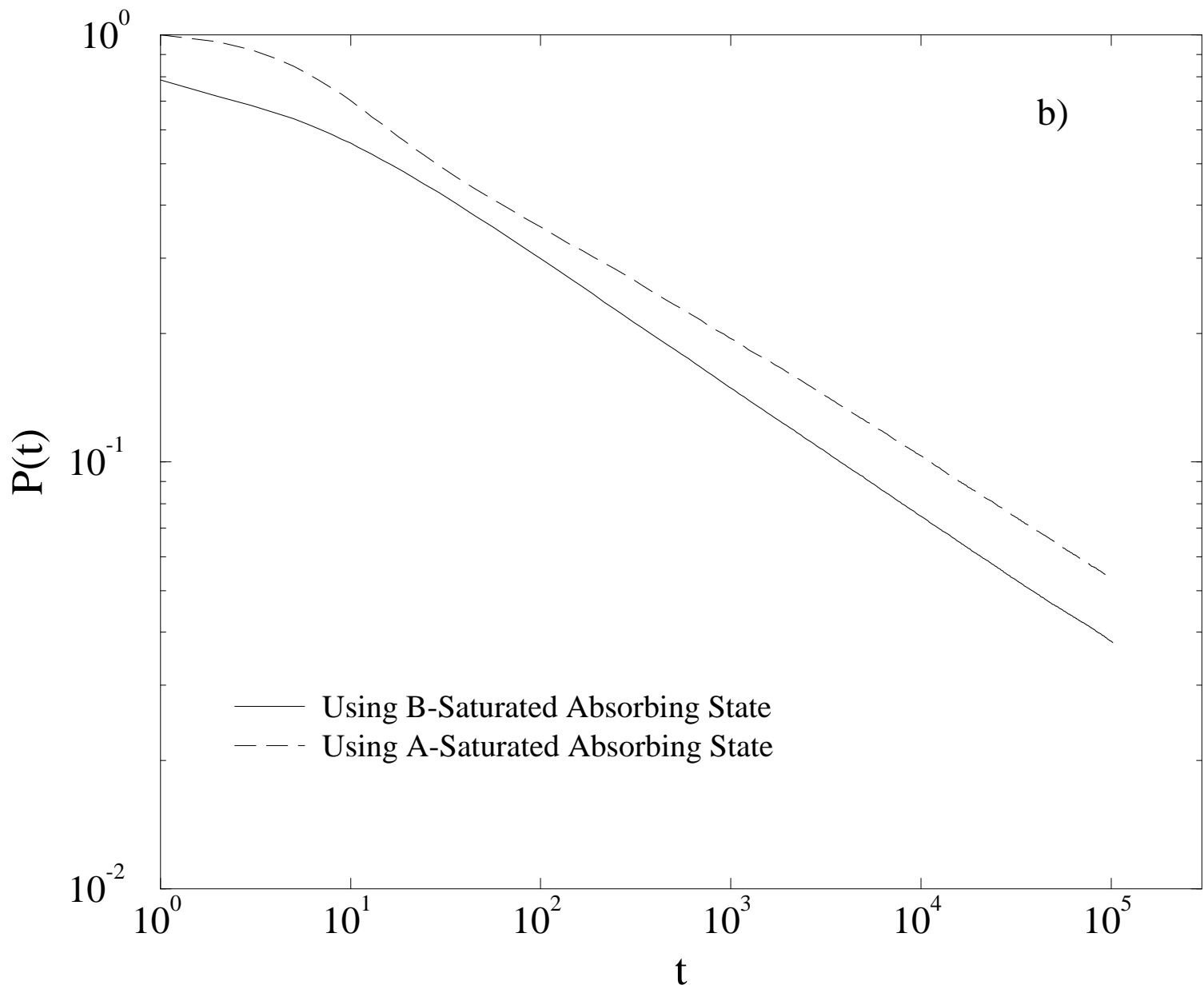
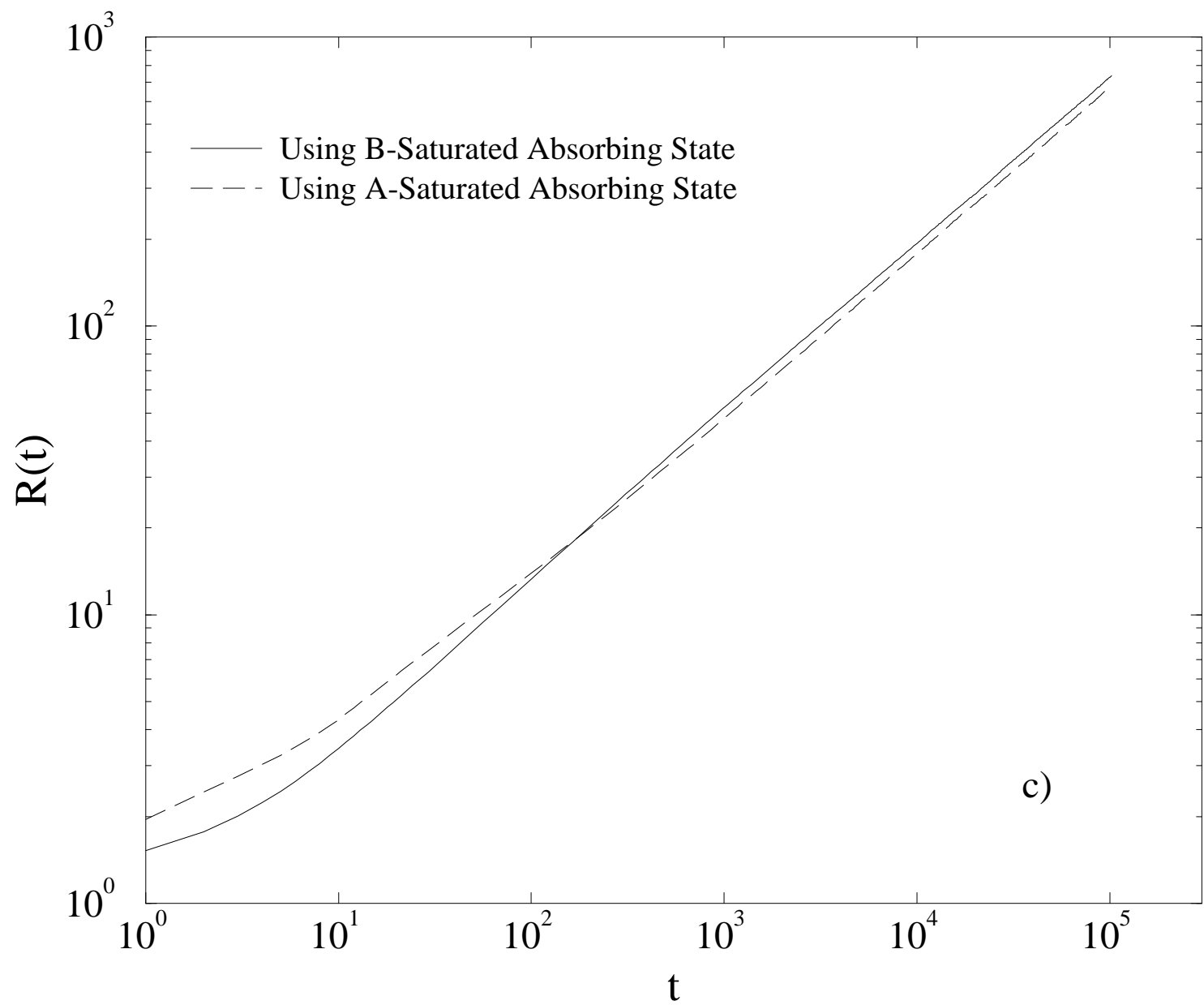
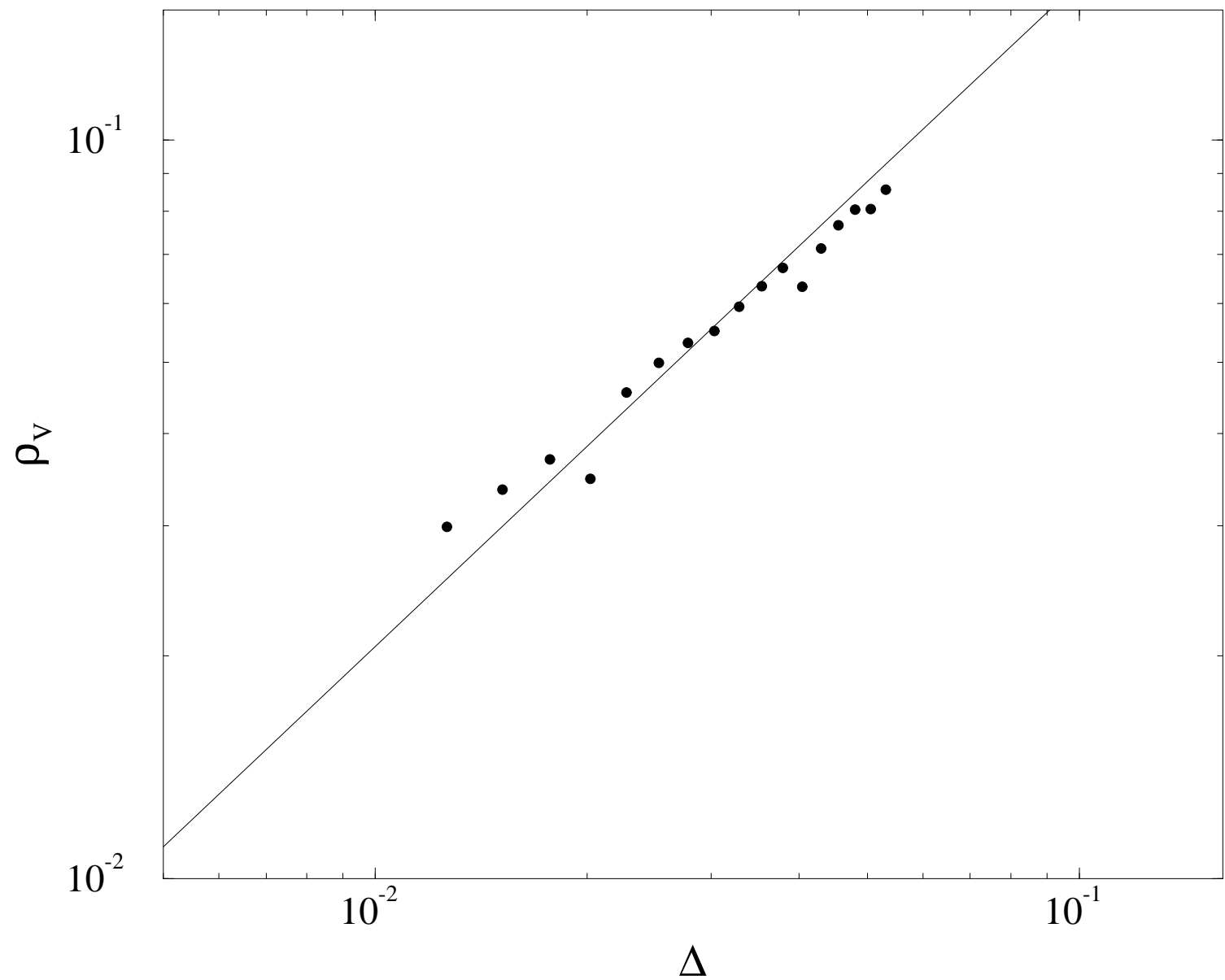
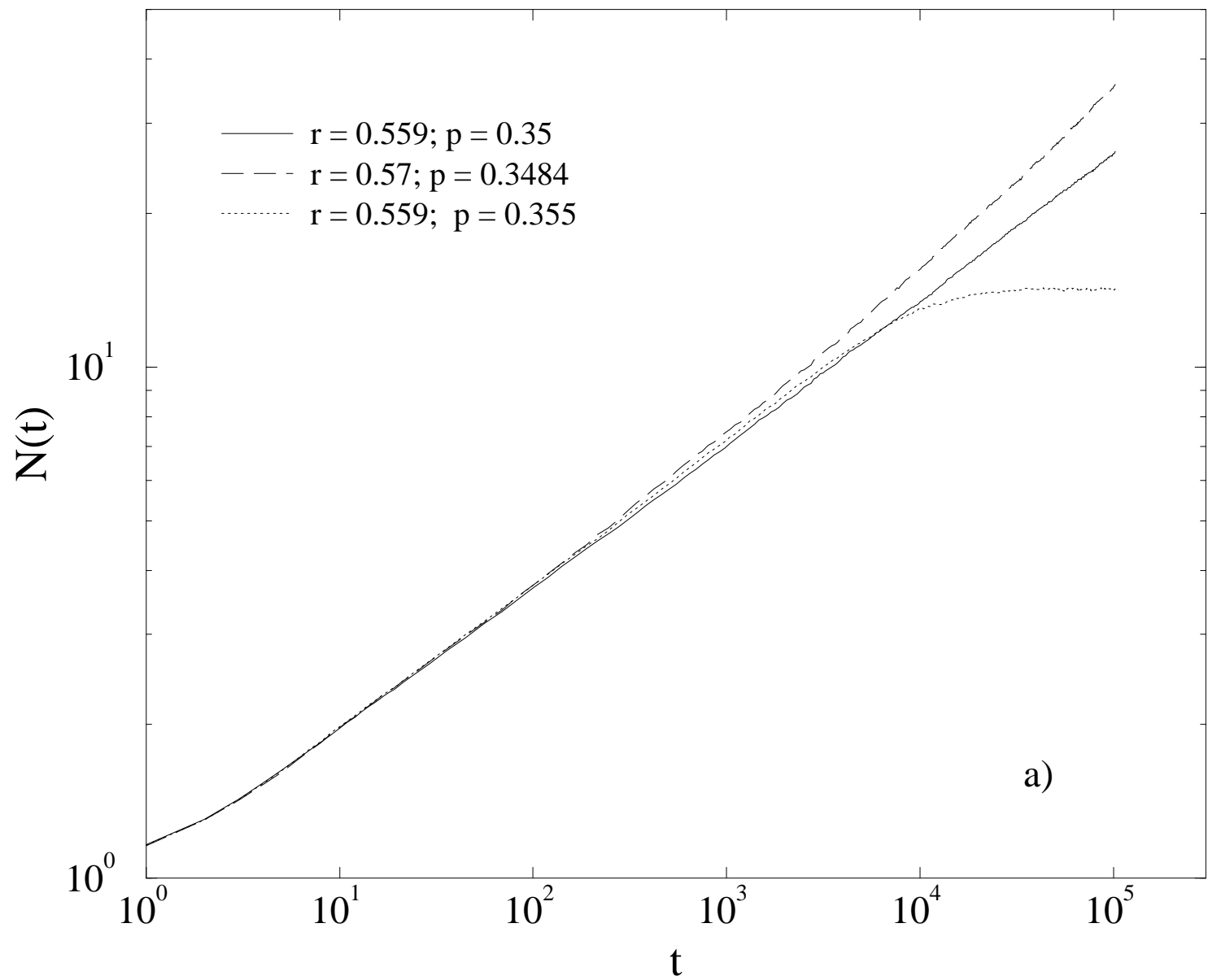
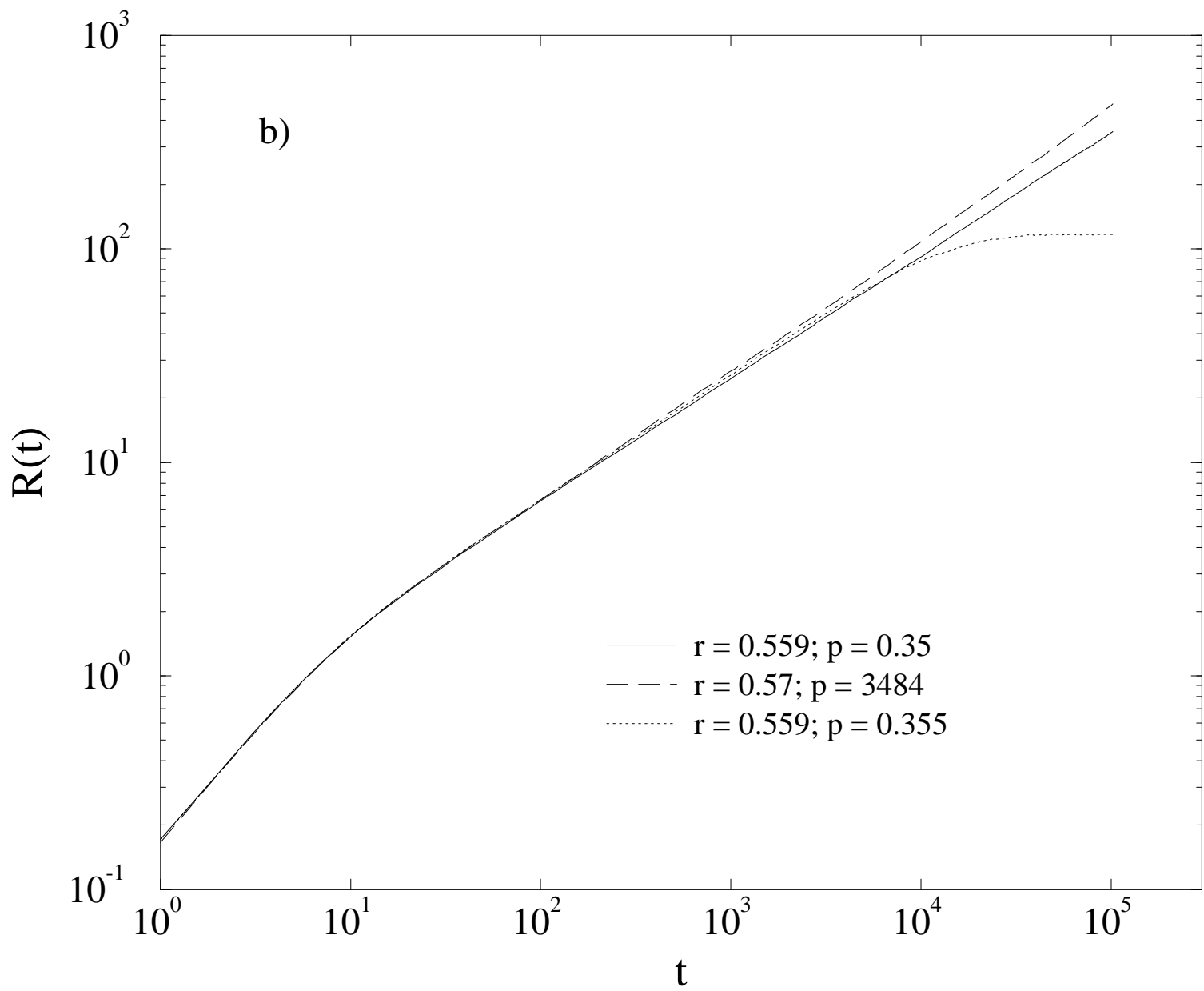
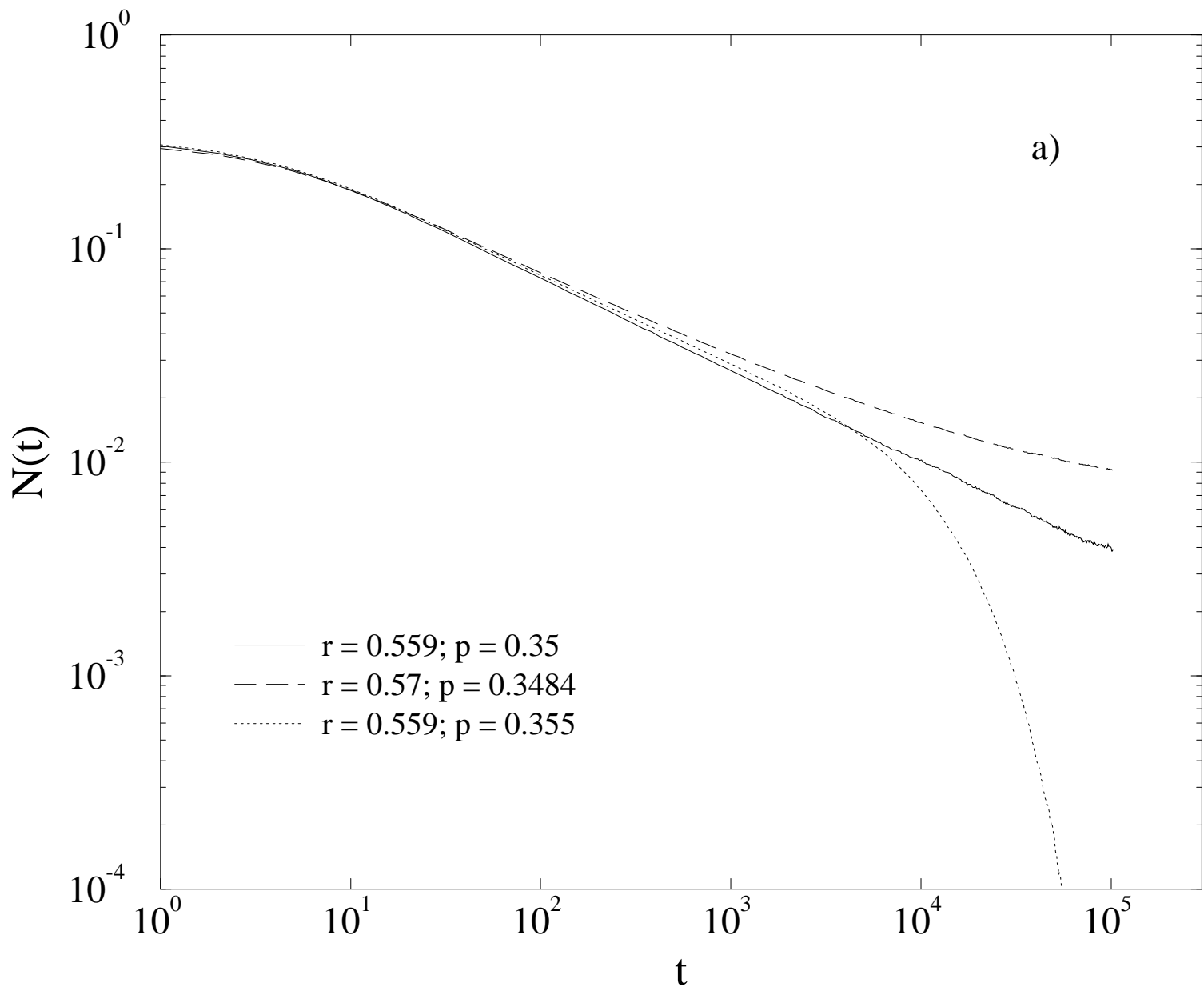
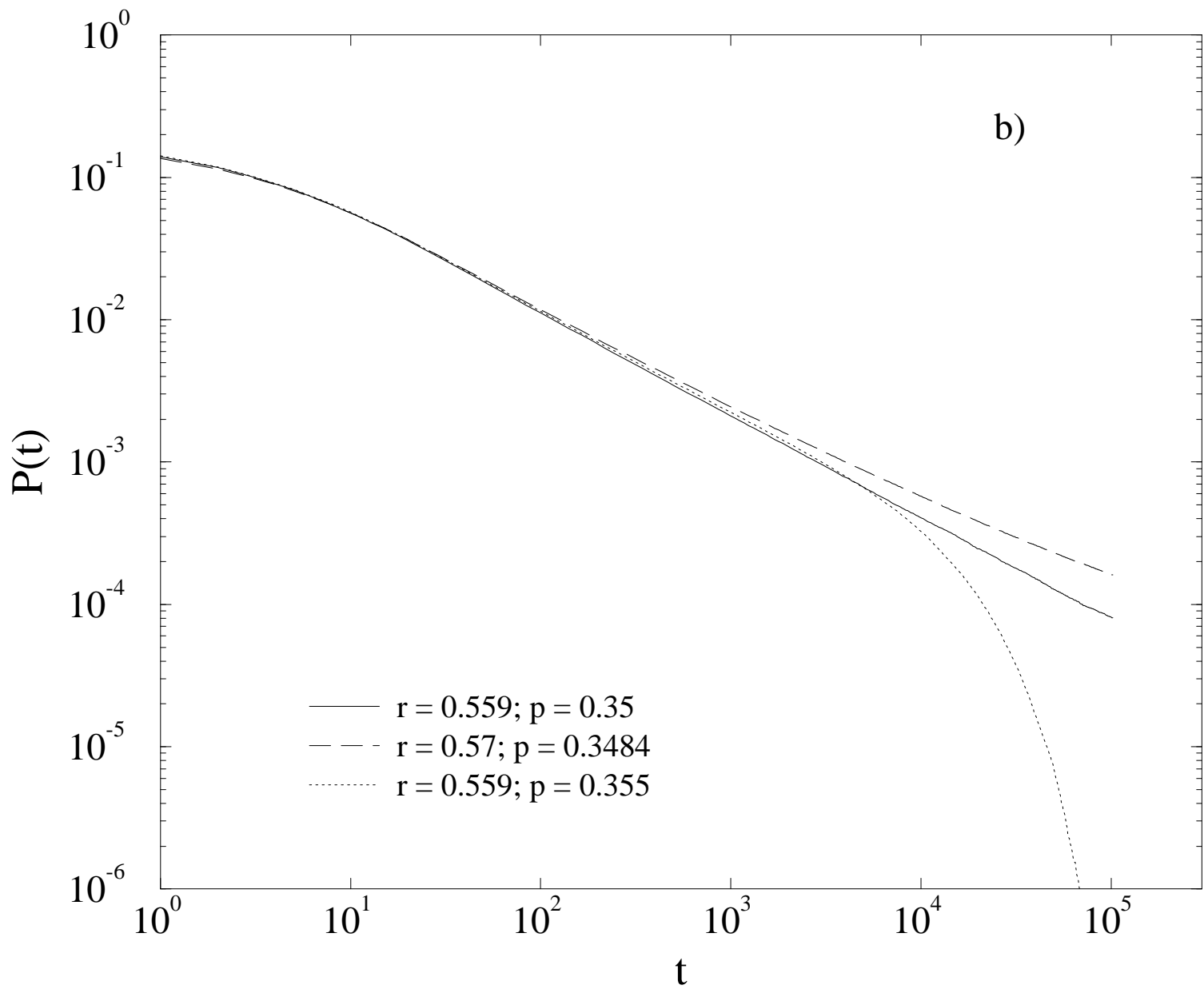
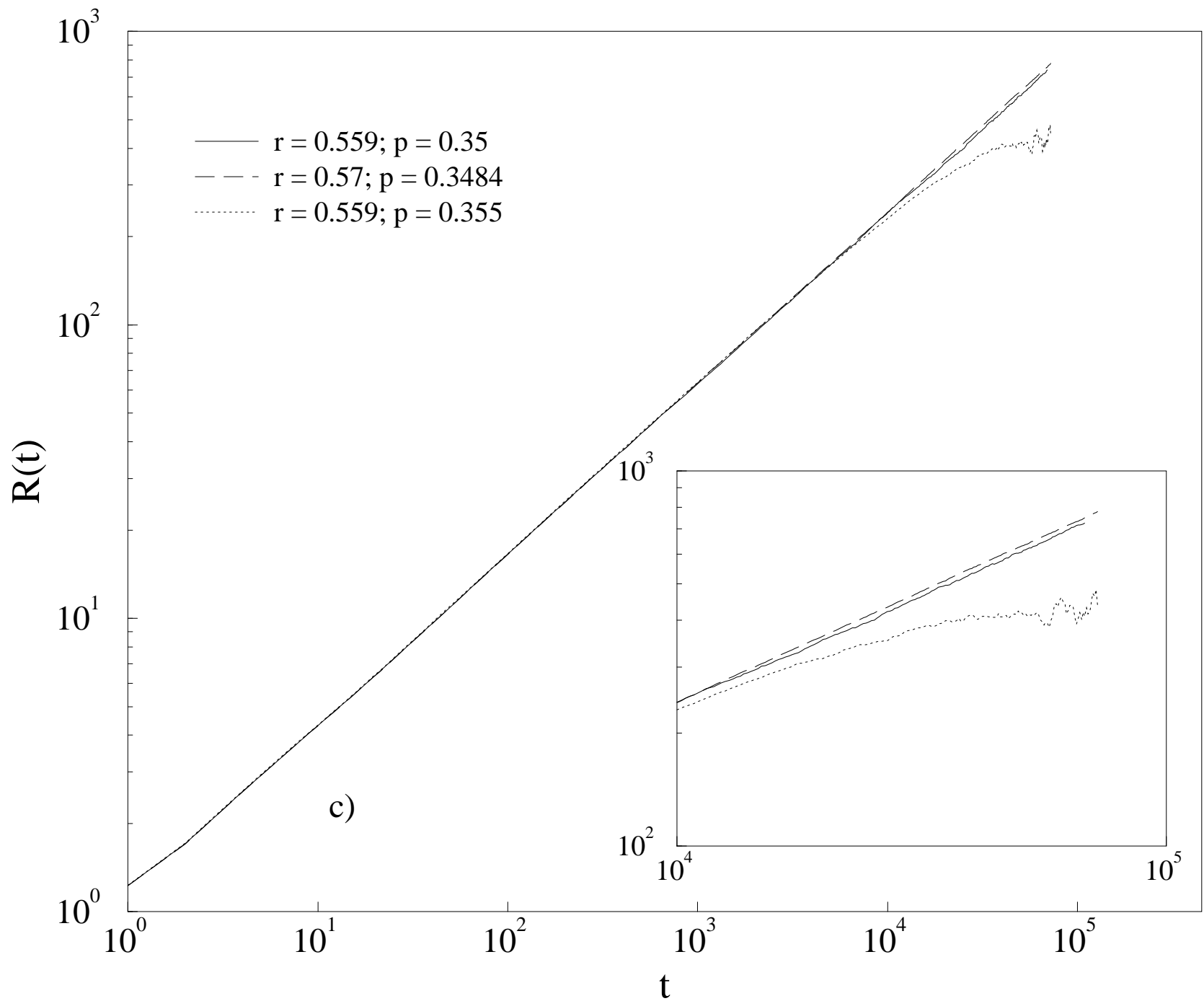
\end{document}